\documentclass[12pt]{article}

\usepackage{sectsty}
\sectionfont{\fontsize{16}{20}\selectfont}
\subsectionfont{\fontsize{14}{16}\selectfont}
\usepackage[letterpaper,margin=1.3in]{geometry}

\usepackage[parfill]{parskip}
\usepackage{setspace}
\usepackage[labelfont=bf,font={small,stretch=0.95}]{caption}
\usepackage[super,sort,compress,numbers]{natbib}

\usepackage{hyperref}

\usepackage{amsmath}
\usepackage{amssymb}
\usepackage{graphicx}
\usepackage{mathptmx}
\usepackage{multirow}

\newcommand{\Loss}{{\fontfamily{cmr}\selectfont \mathcal{L}}}

\usepackage{xcolor}
\usepackage{soul}

\usepackage[version=4]{mhchem}

\begin{document}

\begin{center}

\vspace*{2mm}

{\fontsize{20}{24}\selectfont \bf The structure and topology of an amorphous metal--organic framework \par}

\vspace{4mm}

{\fontsize{14}{18}\selectfont Thomas C. Nicholas,$^{1,\dag}$ Daniel F. Thomas du Toit,$^{1}$ Louise A. M. Rosset,$^{1}$ Davide M. Proserpio,$^{2}$ 
Andrew L. Goodwin,$^{1,\ast}$ and Volker L. Deringer$^{1,\ast}$ \par}

\vspace{4mm}

$^{1}${\em Inorganic Chemistry Laboratory, Department of Chemistry,  University of Oxford,
Oxford OX1 3QR, UK}

\vspace{2mm}

$^{2}${\em Dipartimento di Chimica, Universit\`a{} degli Studi di Milano, Via Golgi 19, Milano 20133, Italy}

\vspace{2mm}

$^\dag$
Present address: {\em Center for Molecular Modeling (CMM), Ghent University, Technologiepark-Zwijnaarde 46, 9052 Zwijnaarde, Belgium}

\vspace{2mm}

$^\ast$\href{mailto:andrew.goodwin@chem.ox.ac.uk}{andrew.goodwin@chem.ox.ac.uk}, \href{mailto:volker.deringer@chem.ox.ac.uk}{volker.deringer@chem.ox.ac.uk}

\end{center}

\vspace{6mm}
\setstretch{1.5}
{\bf 
Amorphous metal--organic frameworks are an important emerging materials class that combine the attractive physical properties of the amorphous state with the versatility of metal--organic framework (MOF) chemistry. The structures of amorphous MOFs have largely been inferred by drawing analogies to crystalline polymorphs and inorganic glasses, but ultimately the validity of such structural models has been challenging to establish either experimentally or computationally. Here we use a unified data-driven approach, combining experimental scattering data and active machine learning for interatomic potentials, to determine the structure of an amorphous zeolitic imidazolate framework (\textit{a}-ZIF)---the canonical amorphous MOF. Our results reveal clear differences between the structure of \textit{a}-ZIF and that of other amorphous tetrahedral networks, allowing us to invalidate the long-standing assumption that these inorganic and hybrid glasses are topologically equivalent. To this end, we introduce a systematic notation for the network topology of amorphous solids, building a bridge to the successful use of topology analysis in crystalline MOFs and to materials informatics. Our work provides insights into the structure and topology of the archetypal amorphous MOF and opens up new avenues for modelling and understanding amorphous framework materials more generally.
}

\section*{Introduction}

Zeolitic imidazolate frameworks (ZIFs) are assembled from tetrahedrally-coordinated transition-metal cations (`nodes') bridged by molecular anions (`linkers'). This hybrid metal--organic architecture resembles the assembly of inorganic zeolite frameworks from Si--O--Si motifs.\cite{Park2006} Both ZIFs and zeolites adopt three-dimensional {\sf AB}$_2$ network structures with diverse topological connectivities, tuneable by choice of composition, synthesis route, and inclusion of structure-directing agents.\cite{Park2006,Zheng2023} Like inorganic zeolites,\cite{Greaves2003,Greaves2007,Masai2023} ZIFs amorphise under heat or pressure;\cite{Bennett2014} the archetypal such system, amorphous zinc(II) imidazolate, is known as `{\it a}-ZIF' and constitutes the metal--organic analogue of silica glass. Amorphous ZIFs are of particular current interest for practical applications: for example, their irregular network connectivity gives rise to highly moldable, isotropic structures without the grain boundaries that would be found in the crystalline ZIF counterparts, and this structural flexibility can be exploited in the design of protective coatings or display technologies.\cite{Bennett2014,Shichun2019,Ding2024,Bennett2024}

The network structures of glasses and related amorphous materials are key to rationalising their physical properties, and yet there is no definitive structural model for \textit{a}-ZIF to date. From an experimental viewpoint, the difficulty is in the relative information poverty of X-ray and neutron total scattering and solid-state NMR data, which are sensitive to local bonding arrangements but cannot reveal the longer-range network topology. The experimental data of ref.~\citenum{Bennett2010} were interpreted in terms of a continuous random network (CRN) structure derived from historic models of amorphous Si, but the actual relevance of this model to \textit{a}-ZIF is entirely unknown. Likewise, from a computational perspective, the key challenge is creating a `correct' structural model: small-scale density-functional theory (DFT) optimisation,\cite{Adhikari2016} \emph{ab initio} molecular dynamics (MD) simulations,\cite{Gaillac2017} and, more recently, machine-learning-based interatomic potentials (MLIPs)\cite{Castel2024,Du2024,Du2024b,Shi2024,Yuan2024} have been used to generate structural models of $a$-ZIF, but achieving fully quantitative agreement with experimental data remains a challenge.

A powerful methodology for determining the structure of amorphous materials that combines experimental measurements and interatomic potentials is the hybrid reverse Monte Carlo (HRMC) approach. HRMC refinements optimise atomic coordinates in large structural models by jointly optimising the quality of fit-to-data and the system's energetics, encoded in a joint loss function.\cite{Opletal2022} Historically, HRMC refinements relied on relatively simple interatomic potentials by virtue of the computational expense involved, but recent advances have allowed the extension to state-of-the-art potentials, \emph{e.g.}, as employed in recent studies of amorphous calcium carbonate\cite{Nicholas2024} or oxygenated amorphous carbon.\cite{Zarrouk2024} In the present work, we use the HRMC paradigm to drive structural exploration and iterative active learning of MLIPs for \textit{a}-ZIF, analogous to recent work for \ce{CeO2} reported in ref.~\citenum{Cuillier2024}, but now moving to the distinct structural and chemical complexity inherent to metal–organic hybrid materials. We show that active learning turns out to be important in the case of \textit{a}-ZIF, because HRMC probes regions of the configurational space that would not be fully covered by MD-driven exploration (which, in itself, has of course been used for very many materials, including silica,\cite{Erhard2024} the silica--water system, \cite{Roy2024} and indeed \textit{a}-ZIF \cite{Castel2024, Du2024, Shi2024}).

Anticipating the results to come, our paper begins with a validation of the active-learning HRMC (AL-HRMC) method used in our study. We then report the structure of \textit{a}-ZIF itself, as determined by applying this method to the original neutron and X-ray total scattering data of ref.~\citenum{Bennett2010}. We compare the local geometric features of our structural model with those of representative crystalline ZIF phases, and subsequently compare the topological features (\emph{e.g.},\ ring statistics) with those of other amorphous tetrahedral networks: namely, the canonical inorganic analogues of \textit{a}-ZIF, amorphous silica and elemental silicon. We show that the ring-size distributions evolve systematically with increasing structural and chemical complexity, highlighting how the greater flexibility and compositional diversity of \textit{a}-ZIF allows access to a broader range of local topological environments compared to these purely inorganic tetrahedral networks. Finally, we quantify these differences by introducing topological labels based on those that are widely employed in the analysis of conventional crystalline network structures.

\section*{Results}

\subsection*{Data-driven modelling of amorphous structures}

\begin{figure}
    \centering
    \includegraphics[width=\linewidth]{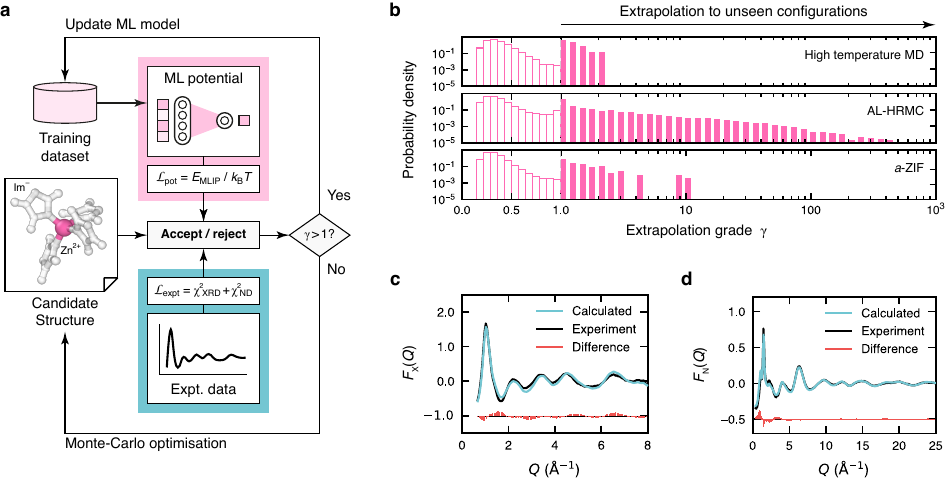}
    \caption{{\bf Data-driven modelling of amorphous structures.} 
    (\textbf{a}) Overview of the approach. Our HRMC refinements minimise a loss function comprising two terms: the difference between experimental and computed data ($\mathcal{L}_{\rm expt}$, turquoise) and the ML-predicted potential energy ($\mathcal{L}_{\rm pot}$, pink). Moves are accepted or rejected based on the Metropolis--Hastings criterion. By calculating the uncertainty of the ML prediction, denoted $\gamma$,\cite{Lysogorskiy2023} for HRMC configurations, we identify structures for evaluation and inclusion in the training set.
    (\textbf{b}) Uncertainty quantification. We distinguish between `interpolative' and `extrapolative' behaviour using a threshold of $\gamma=1$,\cite{Lysogorskiy2023} and show separately normalised probability density values over linear and logarithmic grids (open and filled bars), respectively. While high-temperature (1,500\,K) MD simulations starting from crystalline structures remained mostly interpolative, HRMC explores distinct regions of configurational space, including some atomic environments with $\gamma \gg 10$ (the middle panel characterises structures that were added to the training dataset). The most extrapolative atomic environments in our final \textit{a}-ZIF models have $\gamma \approx 10$.
    (\textbf{c}) X-ray and (\textbf{d}) neutron structure factors from experiments (taken from ref.~\citenum{Bennett2010}) and our \textit{a}-ZIF HRMC refinements (`Calculated'), showing that our final structural model reproduces the experimental total-scattering data from both sources well.
    }
    \label{fig1}
\end{figure}

We started by performing a conventional HRMC loop for \textit{a}-ZIF---with the energy computed using a custom-fitted Atomic Cluster Expansion (ACE)-based MLIP\cite{Drautz2019,Lysogorskiy2021,Bochkarev2022}---during which individual atom moves were proposed, the resulting changes in goodness-of-fit and energy  calculated, and then each move accepted or rejected according to the usual Metropolis criterion (Fig.~\ref{fig1}a). At regular intervals in this process, the HRMC configurations were analysed using the ACE extrapolation grade,\cite{Lysogorskiy2023} $\gamma$, to quantify the extent to which the MLIP model was being forced to extrapolate beyond the specific configurations encountered during training (\emph{i.e.}, $\gamma\gg1$). As Fig.~\ref{fig1}b shows, the requirement of fitting to experiment exposed the HRMC configurations to geometries that were not readily accessed in conventional melt--quench MD protocols. By selecting configurations of sufficiently large $\gamma$ and labelling these with quantum-mechanical data, we expanded the training dataset for the next generation of potential fitting---in analogy to established iterative training protocols, \cite{Deringer2021} and to ref.~\citenum{Cuillier2024} for HRMC. This active-learning protocol (Fig.~\ref{fig1}a) was applied over three iterations, resulting in an MLIP model for \textit{a}-ZIF that was stable in both MD simulations and HRMC refinements. Supplementary Note 1 provides full details of the dataset generation and MLIP fitting. Our final step was then to use this converged potential---first in melt-quench MD simulations, followed by a series of HRMC refinements---with much larger structural models (see Methods section) to arrive at a set of representative \textit{a}-ZIF configurations which formed the basis for our subsequent analysis.

The structure factors calculated for these \textit{a}-ZIF models (Fig.~\ref{fig1}c and d) are essentially as good as those obtained using reverse Monte Carlo (RMC) and reported in ref.~\citenum{Bennett2010}---this is remarkable because RMC is free to fit the data without any constraints due to energetics, and therefore it is expected to reproduce the scattering data on its own very well. In contrast, our HRMC $a$-ZIF models are at once directly based on experimental data and on accurate energetics from an MLIP: they are more stable than the corresponding RMC model by about 250 kJ mol$^{-1}$ (see Supplementary Note 3 for a discussion of similarities and differences between the two types of structural models). Whereas the RMC models of ref.~\citenum{Bennett2010} were generated by enforcing network connectivity inherited from an earlier model of amorphous silicon, the MD melt-quenching stage of our simulation protocol removes the need for assuming a pre-defined topology for \textit{a}-ZIF. Instead, it allows the topology to emerge naturally from network rearrangements, as directed by our ML potential.

\subsection*{Local and intermediate-range order in \textit{a}-ZIF}

\begin{figure}
    \centering
    \includegraphics[width=\linewidth]{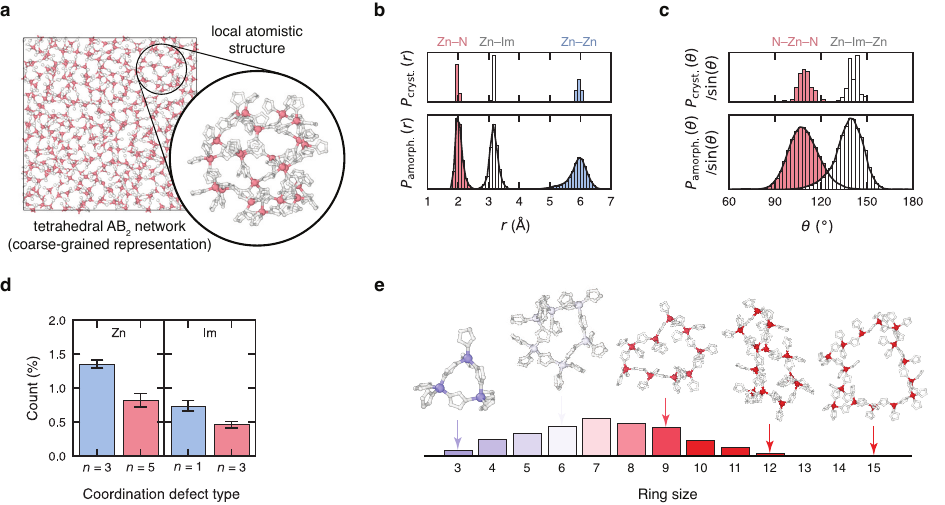}
    \caption{{\bf The structure of \textit{a}-ZIF.}
    (\textbf{a}) A representative slice of our structural model, with imidazolate linkers coarse-grained to a single bead positioned at the geometric centroid of each ring, highlighting the underlying tetrahedral {\sf AB}$_2$ network. The inset shows a representative atomistic region of the structural model.
    Distributions of (\textbf{b}) bond lengths and (\textbf{c}) bond angles in crystalline ZIFs (top) and in our \textit{a}-ZIF structural model (bottom).
    (\textbf{d}) Coordination defect statistics for zinc centres (left) and imidazolate linkers (right), with whiskers representing the standard error of the mean across the five \textit{a}-ZIF models.
    (\textbf{e}) Distribution of ring sizes in the \textit{a}-ZIF model, illustrated with example ring fragments extracted from the model, spanning from the smallest 3-membered rings to the largest 15-membered rings.
    Structures were visualised using OVITO.\cite{Stukowski2010} 
    Hydrogen atoms have been omitted for visual clarity.
    }
    \label{fig2}
\end{figure}

The \textit{a}-ZIF model exhibits an extended tetrahedral {\sf AB}$_2$ network structure with chemically sensible local coordination geometries, as illustrated by a representative coarse-grained slice shown in Fig.~\ref{fig2}a (with the inset highlighting local coordination). This point is made clear in the pair and angle correlation functions: in Fig.~\ref{fig2}b--c, we compare some key distributions obtained from our model with those for crystalline ZIFs. Naturally, these distributions are broader in the amorphous structures than in the crystalline polymorphs. The {\sf AB}$_2$ network topology was not fixed in our HRMC refinements: therefore, the system was free to incorporate coordination defects should they be energetically allowed and consistent with experiment. We do indeed observe a small fraction of coordination defects, comprising $1.4~\%$ (3-coordinate) and $0.8~\%$ (5-coordinate) zinc nodes (Fig.~\ref{fig2}d), consistent with experimental solid-state $^{15}$N NMR measurements of amorphous ZIFs,\cite{Baxter2015,Bode2025} and with previous ML-driven MD simulations.\cite{Castel2024,Yuan2024} Coordination defects are approximately twice as prevalent for zinc centres compared to imidazolate linkers. Notably, under-coordinated (trigonal) and over-coordinated (trigonal-bipyramidal) zinc geometries have been implicated in the temperature-induced reconstructive phase transition between the \textbf{coi} and \textbf{zni} ZIF topologies.\cite{Schroder2013}

Intermediate-range order in \textit{a}-ZIF is normally quantified in terms of the distribution of network ring sizes, which has been a point of discussion amongst earlier studies.\cite{Bennett2010,Adhikari2016,Gaillac2020,Castel2024} Our HRMC-derived \textit{a}-ZIF model exhibits a broad, unimodal ring distribution spanning sizes from three to about 15 zinc nodes (Fig.~\ref{fig2}e). This diversity of ring sizes is consistent with the ring statistics of crystalline ZIF polymorphs, albeit the crystalline state tends empirically to favour even-membered rings (perhaps a result of crystal symmetries). Although 3-membered rings are absent in crystalline ZIFs, we argue that their presence in our amorphous ZIF model is reasonable, given the modest energetic penalty relative to 4-membered rings and their occurrence in other disordered networks such as \textit{a}-Si and \textit{a}-\ce{SiO2} (Supplementary Note 5). In terms of consistency with experiment, we mention two observations. First, in principle, $^{67}$Zn solid-state NMR measurements offer some sensitivity to Zn environments in different rings, and so the broad unimodal distribution in Fig.~\ref{fig2}e is consistent with the experimental observation of a similarly broad, unimodal, and asymmetric distribution of Zn chemical shifts in \textit{a}-ZIF.\cite{Madsen2020} And, second, the ring statistics of glass-forming network structures are related to the so-called `fragility' of the resulting amorphous state;\cite{Shi2023} consequently, the relatively large ring sizes we observe for \textit{a}-ZIF are consistent with the observation that ZIF glasses have low experimental fragilities.\cite{Bennett2015}

\subsection*{Relationship to other tetrahedral random networks}

\begin{figure}
    \centering
    \includegraphics[width=\linewidth]{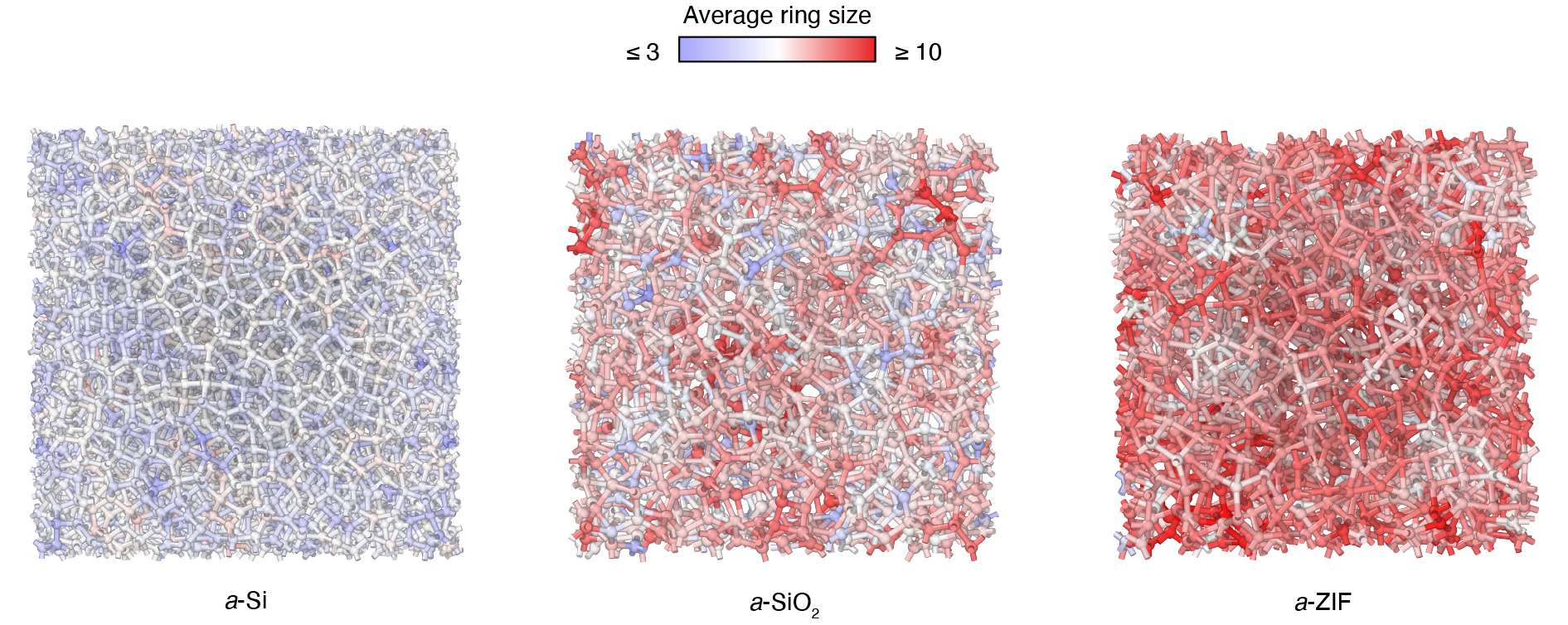}
    \caption{
        {\bf Differences in the topologies of amorphous tetrahedral networks.}
        We compare structural models of amorphous silicon (left),\cite{Deringer2018} amorphous silica (middle),\cite{Erhard2022} and \textit{a}-ZIF (right), the latter two being shown as coarse-grained representations (only the Si and Zn nodes are shown, respectively). Nodes are colour-coded by the average size of rings in which they are found, with blue indicating smaller average ring sizes and red indicating larger ones.
        Structures were visualised using OVITO.\cite{Stukowski2010}
    }
    \label{fig3}
\end{figure}

We turn now to the unresolved question of the similarity of the network topology of \textit{a}-ZIF to other tetrahedral CRN structures. We compare our $a$-ZIF structure to the high-quality, MLIP-derived structural models of amorphous silicon and amorphous silica published in refs.~\citenum{Deringer2018} and \citenum{Erhard2022}, respectively. In Fig.~\ref{fig3}, we make the simplest visual comparison of intermediate-range order in these three systems by illustrating the spatial distribution of rings of various sizes. The point of \textit{a}-ZIF containing particularly large rings is made all the more obvious in this representation. As the characteristic length scale of the CRNs---\emph{i.e.}\ the inter-node separation---increases from \textit{a}-Si to \textit{a}-SiO$_2$ to \textit{a}-ZIF, so too do both the fraction of large rings and the diversity of ring sizes. Hence the conceptual link amongst the three systems that involves mapping Zn--imidazolate--Zn linkages onto Si--O--Si and, in turn, onto Si--Si connections may have physical meaning on relatively short length scales, but should not be taken to imply a direct correspondence amongst the resulting network topologies. Indeed, coarse-graining of the underlying network topology in \textit{a}-ZIF reveals a tetrahedral CRN whose internal angles are substantially more varied (further away from 109.5$^{\circ}$ than they are in \emph{a}-Si; see Supplementary Figure 7 for a comparison of local geometric properties).

\subsection*{Quantifying topology}

The observation that three different amorphous materials have CRNs with meaningfully different topological properties prompts the question of how one might label amorphous network structures beyond the use of qualitative terms, such as low- and high-density amorphous (LDA and HDA, respectively) for silicon or water. Amongst the glass community, the focus has historically been on capturing the speciation of node connectivity ($Q^n$ labels\cite{Kohara2021}), but all three systems we compare here are almost entirely four-connected network structures. By contrast, the metal--organic framework community is used to differentiating amongst network topologies even with identical node connectivity using numerical labels of various kinds. The topologies of \textbf{cag}- and \textbf{zni}-ZIF structures, to take a relevant example, can be discriminated by the corresponding vertex symbols $[4.6_2.6.6.6.6]$ and $[4.6.6.6_3.6_2.12_{40}]$, respectively, despite both involving four-connected zinc nodes and two-connected imidazolate linkers.\cite{Friedrichs2005,Blatov2010} Here, the components of each vertex symbol denote the size of the shortest ring at each angle of which the node is a member, with the subscript denoting the number of such rings (omitted if unity).

\begin{figure}
    \centering
    \includegraphics[width=\linewidth]{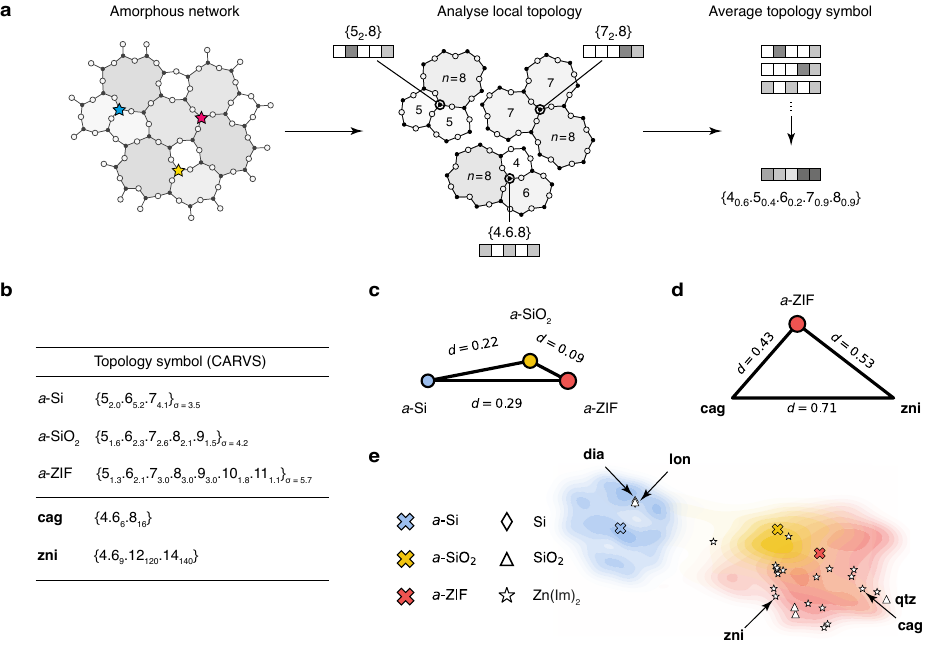}
    \caption{
        {\bf A quantitative measure for topology in amorphous materials.} 
        (\textbf{a}) Schematic illustrating how the topological label for amorphous materials is obtained, using a two-dimensional example (obtained by digitising a CRN sketch from ref.~\citenum{Zachariasen1932}).
        (\textbf{b}) Topology symbols (CARVS) for \textit{a}-Si, \textit{a}-SiO$_2$, and \textit{a}-ZIF. For uninodal networks, such as the crystalline topologies \textbf{cag} and \textbf{zni}, there is only one type of node, resulting in $\sigma = 0$, which is consequently omitted from the topology symbol.
        A topological distance can be calculated between any two topology symbols. Distances between (\textbf{c}) \textit{a}-Si, \textit{a}-SiO$_2$, and \textit{a}-ZIF, and between (\textbf{d}) \textbf{cag}, \textbf{zni}, and \textit{a}-ZIF, are visualised in the style of ref.~\citenum{Nicholas2020}. The size of the node indicates the $\sigma$ value for that configuration.
        (\textbf{e}) A two-dimensional embedding of the topology space defined by \textit{a}-Si (blue), \textit{a}-SiO$_2$ (yellow), and \textit{a}-ZIF (red). The embedding was created by characterising each node with the CARVS-based vectors followed by dimensionality reduction using UMAP,\cite{Mcinnes2018} with the point distribution shown as a heat map. In addition, per-structure amorphous CARVS, along with the per-node crystalline Si (diamonds), silicate (triangles), and ZIF (stars) CARVS were projected onto the same embedding.
    }
    \label{fig4}
\end{figure}

In this spirit, we extend the vertex symbol formalism to amorphous network structures by introducing several key modifications. Instead of considering only the smallest rings around each node, we account for all ring sizes, arranging them in order of increasing size. For each ring size, we calculate the average number of such rings per node across the network, which results in fractional indices (Fig.~\ref{fig4}a). To simplify notation, any ring sizes with an average count of less than one per node are omitted. We refer to this symbol as the `Cumulative All-Rings Vertex Symbol' (CARVS) (see Supplementary Note 4 for further details on its calculation and a comparison with the vertex symbol for crystalline topologies, and note the conceptual similarity to Schl\"afli cluster analysis \cite{Treacy2000,Hobbs2008}). The resulting CARVS topology symbols for \textit{a}-Si, \textit{a}-SiO$_2$, and \textit{a}-ZIF are presented in Fig.~\ref{fig4}b, with the final subscript value $\sigma$ quantifying the standard deviation amongst node environments. An advantage of these representations is that the progression from smaller to larger rings and the increased diversity of ring sizes with increasing characteristic linker length are both immediately apparent. Furthermore, the measure
\begin{equation}
d(\alpha,\beta)=\frac{1}{\sqrt{2}}\left|\tilde{\mathbf r}(\alpha)-\tilde{\mathbf r}(\beta)\right|,
\end{equation}
formed using the relative topological vectors
\begin{equation}
\tilde{\mathbf r}=\frac{1}{\sum_ia_i}[a_1,a_2,\ldots a_n]
\end{equation}
based on ring counts $a_i$ (up to some maximum size $n$), is a distance metric that quantifies the topological dissimilarity between structures $\alpha$ and $\beta$. By construction, $0\leq d\leq1$ with $d=0$ implying identical (relative) ring statistics and $d=1$ denoting entirely dissimilar statistics. The (dis-) similarities between amorphous CRN structures, and between \textit{a}-ZIF and the closely-related crystalline ZIF polymorphs ZIF-\textbf{cag} and ZIF-\textbf{zni} are visualised in Fig.~\ref{fig4}c and \ref{fig4}d, respectively. What emerges naturally from this analysis is that the two {\sf AB}$_2$ CRNs are more topologically similar to one another than either is to \textit{a}-Si, and that \textit{a}-ZIF is topologically intermediate to the two crystalline phases, \textbf{cag}- and \textbf{zni}-ZIF, that bookend its thermal stability field.\cite{Bennett2010}

Finally, we note that developing suitable descriptors is a critical step in materials informatics and `design',\cite{Ramprasad2017} and our proposed topology symbol approach may help to extend this paradigm from crystalline to amorphous systems, the latter being very much still at an early stage.\cite{Liu2025} Indeed, it was recently shown that a descriptor capturing the heterogeneity of the local topology can predict the conductivity differences between various classes of carbon structures.\cite{Iwanowski2025} Figure~\ref{fig4}e illustrates how a unified amorphous--crystalline topological space can be constructed using the CARVS vectors. Here, the per-node amorphous CARVS are first embedded using UMAP,\cite{Mcinnes2018} with their point distribution visualised as a heat map. The per-node CARVS vectors for the crystalline topologies (unfilled markers), along with the per-structure amorphous CARVS (coloured crosses), are then projected onto this embedding as distinct markers. The topological space spanned by $a$-Si is clearly distinct from those of $a$-\ce{SiO2} and $a$-ZIF, with $a$-Si environments clustering near the \textbf{dia} topology. In contrast, the topological spaces of $a$-\ce{SiO2} and $a$-ZIF overlap, albeit with $a$-ZIF exhibiting a broader distribution.

\section*{Discussion}

Our analysis has succeeded not only in characterising the structure of \textit{a}-ZIF---in as much as it is possible to do so---but it has helped us to place the intermediate-range order found in this structure within the broader context of amorphous and crystalline tetrahedral networks. In the same way that structure--property relationships have been established between (periodic) network topologies and physical properties such as porosity,\cite{Shevchenko2020} we anticipate that the quantification of amorphous network topologies may help to reveal {\em disorder}--property relationships in due course. Indeed, it has been found that the framework topology in MOFs affects both chemical and mechanical properties\cite{Tan2010,Meekel2024}---and so our work points towards the more systematic understanding and control of topology as a strategy for targeted design, not only of crystalline MOFs but ultimately of their amorphous counterparts. For example, it has been recently demonstrated that the enumeration of merged nets---\textit{i.e.,} combinations of simpler crystalline network topologies---offers a systematic approach for predicting and constructing intricate multicomponent MOF platforms.\cite{Jiang2024} Extending this methodology to amorphous materials, which are particularly relevant for hybrid-material applications,\cite{Bennett2015,Bennett2024} requires an analogous, quantitative framework to describe amorphous topologies.

From a methodological viewpoint, our study strengthens the case for the long-term goal of unified structure determination, combining experimental and first-principles data in an active-learning loop, in a case for which MD simulations on their own are insufficient.
Specifically, we arrived at a robust structural model of \textit{a}-ZIF by leveraging the combination of accuracy and raw computational speed afforded by carefully parameterised ACE ML interatomic potentials. Our reliance here on an experimentally-informed active-learning approach---rather than conventional MD-based training---reflects what we perceive to be the particular difficulty of computationally navigating the configurational landscape of network glasses based on molecular components. 
Zarrouk {\em et al.}\ have recently pointed out that not only diffraction data, but also X-ray spectroscopy and other experimental probes may be used in HRMC refinements, \cite{Zarrouk2024} and in the long run one may envision a multi-fidelity, multi-component approach that takes the different probes into account on an appropriate footing.
Our current model has been based on a hand-crafted ZIF potential as a starting point, but beyond this, recent successful applications of fine-tuning foundational interatomic potentials (see, \emph{e.g.}, ref.~\citenum{Kaur2024}) suggest that similar approaches could be used here as well. The general ideas put forward in the present work (and those on which it builds) remain unchanged by the question how exactly the ML interatomic potential is fitted.

Beyond {\em a}-ZIF, we expect that AL-HRMC approaches may prove useful in determining structural models for a variety of network glasses that are otherwise difficult to characterise, including amorphous Prussian blue analogues,\cite{Ma2022} formates,\cite{Wang2004}, dicyanometallates,\cite{Ghalsasi2012} and carboxylate-based MOFs.\cite{Kim2024,Xue2024} As our understanding of the topological diversity of such phases develops, so too will our ability to control functional response through amorphous materials design---for example, by changing the chemical composition or synthesis conditions.\cite{Liu2025} More widely, the notion of topological diversity amongst tetrahedral networks may have relevance well beyond MOF chemistry, and we flag in particular the conceptual parallels to topological transitions in models of hydrogen-bonded liquid structures.\cite{Neophytou2022}

\clearpage

\setstretch{1.1}

\section*{Methods}

{\bf Hybrid reverse Monte Carlo.} We carried out HRMC refinements following the general ideas in ref.~\citenum{Opletal2002}. We propose randomised atomic moves and accept these based on the combined loss function,
\begin{equation}
    \Loss = \Loss_{\rm expt} + \Loss_{\rm pot},
\end{equation}
where the former contribution measures the deviation from experiment, and the second term adds a penalty according to the potential energy of a configuration, here measured by quantum-mechanically accurate ML interatomic potentials. For the fit to experiment, we use
\begin{equation}
    \Loss_{\rm expt} = \chi^{2}_{\rm XRD} + \chi^{2}_{\rm ND},
\end{equation}
that is, a sum of the deviations for X-ray and neutron scattering data. (In the absence of other information, we weight both terms equally in this expression.) The experimental data were those reported in ref.\,\citenum{Bennett2010}. The corresponding total scattering functions were computed from our \textit{a}-ZIF models by weighted Fourier transform of the partial pair correlation functions, $g_{ij}(r)$,\cite{Keen2001} themselves obtained as histograms with a bin width of $0.02$~\AA{}. The second term in the overall loss is
\begin{equation}
    \Loss_{\rm pot} = \frac{E_{\rm MLIP}}{k_{\rm B}T},
\end{equation}
where $E_{\rm MLIP}$ is evaluated using bespoke ML interatomic potentials as described below.

Starting configurations were generated using a multistep protocol. First, \textit{a}-\ce{SiO2} models were generated using the melt-quench protocol described in ref.~\citenum{Erhard2022}, combining the CHIK \cite{Carre2008} and \ce{SiO2}-GAP-22 potentials.\cite{Erhard2022} Systems containing 1,728 formula units were used, consistent with the \textit{a}-\ce{SiO2} study of Ref.~\citenum{Erhard2022}. These structures were then re-scaled such that the final \textit{a}-ZIF models would have the same density as used in ref.~\citenum{Bennett2010}. The \textit{a}-\ce{SiO2} models were transformed (`backmapped') to ZIF models by replacing Si with Zn and O with Im, oriented such that the centre of the ring coincided with the {\sf B} site coordinates, and the $C_2$ symmetry axis of the molecule aligned with the vector bisecting the {\sf A}–{\sf B}–{\sf A} bond angle. The backmapping procedure is described in ref.~\citenum{FaureBeaulieu2023}.

Individual moves were proposed and accepted or rejected according to the usual Metropolis–Hastings criterion. These moves involved a combination of atomic displacements (maximum value 0.1~\AA{}) and rigid-body translations or rotations (maximum values of 0.1~\AA{} and 1.0\textdegree, respectively) of imidazolate molecules.

The AL-HRMC protocol required a relatively modest system size, small enough to be suitable for evaluating energies and forces with DFT, yet large enough to meaningfully represent the local and intermediate-range structural order. We therefore used \textit{a}-ZIF configurations with 32 formula units (544 atoms) in a cubic box with an edge length of $a=19.6$~\AA{}. The \textit{a}-ZIF models created using the final potential contain 1,728 formula units (29,376 atoms) in a cubic box with $a=74.1$~\AA{}.

{\bf ML interatomic potentials.} 
MLIPs were fitted using the Atomic Cluster Expansion (ACE) framework\cite{Drautz2019} as implemented in the \texttt{pacemaker} software.\cite{Bochkarev2022,Lysogorskiy2021} This framework enables very-large-scale, computationally efficient simulations that do not require GPU computing resources at runtime. \cite{Zhou2025}

In brief, ACE MLIP models are built from atomic properties, $\varphi_i$, expanded across body-ordered functions from the set of neighbours of each atom. For all species, the atomic neighbour density was determined up to a radial cut-off of 6.0 \AA{}, and expanded using 600 basis functions per element. 
We describe the atomic energy, $E_i$, with a non-linear embedding consisting of 6 atomic properties, $\varphi_i$:
\begin{equation}
    E_i = \varphi_i^{(1)} + \sqrt{\varphi_i^{(2)}} + \sum_{j=3}^6 (\varphi_i^{(j)})^{f_j},
\end{equation}
where the first two terms correspond to a Finnis--Sinclair-type embedding. \cite{Finnis1984} The exponents of the remaining four atomic properties were determined during hyperparameter optimisation (see ref.~\citenum{ThomasduToit2024} for details), yielding $f_j \in \{ 0.975, 0.635, 0.025, 0.540 \}$. Complete details on dataset generation and the fitting procedure for the MLIP models are provided in Supplementary Note 1.

{\bf Reference data.} DFT computations were performed using VASP,\cite{Kresse1993,Kresse1996b} version 6.4.1. The electron--ion interactions were described using the projector augmented-wave (PAW) method.\cite{Bloechl1994, Kresse1999} The Kohn--Sham equations were solved self-consistently until the total free energy change and the band-structure-energy change converged to within $10^{-8}$~eV per cell. The strongly constrained and appropriately normed (SCAN) meta-GGA functional was used,\cite{Sun2015} which was found to describe the main \ce{Zn(Im)2} polymorphs well (see Supplementary Note 1). The plane wave cut-off energy was 1,000~eV, and partial occupancies were treated using Gaussian smearing with $\sigma=0.2$~eV. Due to the large simulation-cell sizes, reciprocal space was sampled at $\Gamma$ only.

{\bf MD simulations.} MD simulations in the NVT ensemble were carried out using \texttt{LAMMPS},\cite{Thompson2022} using a Nos\'{e}--Hoover thermostat and a temperature damping parameter of 100~fs.\cite{Nose1984,Hoover1985} The timestep was 0.25~fs. The imidazolate molecules were held rigid during MD simulations using the LAMMPS \texttt{rigid/small} fix.

Before the melt-quench MD simulation, the \textit{a}-ZIF structures were equilibrated at 300~K following the `preparation' protocol described in ref.~\citenum{Castel2022}. Initially, the system was relaxed at 20~K for 50~ps with a temperature damping parameter of 40~fs. This was followed by an additional relaxation period at the same temperature for 5~ps with a temperature damping parameter of 10~fs. The system was then heated to 300~K at a rate of 5~K~ps$^{-1}$, with a temperature damping parameter of 100~fs, and finally equilibrated at 300~K for 5~ps.

The melt-quench simulations were conducted in three stages. First, the system was heated to 1,500~K (a temperature sufficiently high to facilitate network reorganisation) over 500~ps. Next, the system was annealed at this temperature for 5~ns, followed by quenching to 600~K at a rate of 10$^{12}$~K~s$^{-1}$. To verify that our interpretation of the $a$-ZIF topology was not strongly influenced by the choice of annealing temperature, additional simulations were performed at lower annealing temperatures (Supplementary Note 2).

{\bf Topology analysis.}
Topology and ring-size analyses were performed using a custom Python interface to the \texttt{CrystalNets.jl} package.\cite{Zoubritzky2022,Zoubritzky2022b} The latter computes rings up to a specified maximum size, taking the edges of the atomic graph as an input. For ZIF structures, the graph was obtained by first assigning Zn--Im neighbours using a Zn--N distance cut-off of $2.5$~\AA{},\cite{Gaillac2017,Castel2024} and then connecting zinc nodes bridged by a common imidazolate linker. For \textit{a}-\ce{SiO2}, the Si--O connectivity was determined with a cut-off of $2.0$~\AA{}, as in ref.~\citenum{Erhard2022}, and silicon nodes connected by a common oxygen atom were linked. For \textit{a}-Si models, a Si--Si distance cut-off of $2.85$~\AA{} was used.\cite{Deringer2018} The calculations and derived per-node topology symbols were cross-checked against point symbol and vertex symbol outputs from the well-established \texttt{ToposPro} software.\cite{Blatov2014}

\section*{Data availability}

Data supporting the present work, including structural models and ACE model parameters, are openly available at
\url{https://github.com/tcnicholas/amorphous-mof}.

\section*{Code availability}

The code for coarse-graining and backmapping of ZIF structures is openly available at \url{https://github.com/tcnicholas/chic}.
The HRMC implementation used and the structure analysis scripts are openly available at \url{https://github.com/tcnicholas/amorphous-mof}. 
The code for performing rings and topology analysis is openly available at \url{https://github.com/tcnicholas/topo-metrics}. 
Other codes were used as provided by their respective authors.

\section*{Acknowledgements}

We thank Linus Erhard and Igor Baburin for valuable discussions.
T.C.N. was supported through an Engineering and Physical Sciences Research Council DTP award [grant number EP/T517811/1]. 
D.M.P. thanks the MUR for the grant PRIN2020 ``Nature Inspired Crystal Engineering
(NICE)'' and Prof.\ Vladislav A. Blatov at the Samara Center for Theoretical Materials Science for providing the free ToposPro software.
A.L.G. acknowledges financial support from the ERC (grant no.\ 788144).
V.L.D. acknowledges a UK Research and Innovation Frontier Research grant [grant number EP/X016188/1].
The authors would like to acknowledge the use of the University of Oxford Advanced Research Computing (ARC) facility in carrying out this work (https://doi.org/10.5281/zenodo.22558).
We are grateful for computational support from the UK national high performance computing service, ARCHER2, for which access was obtained via the UKCP consortium and funded by EPSRC grant ref EP/X035891/1.

\end{document}


\begin{center}
\noindent
{\Large \textbf{Supplementary Information for}\\
`The structure and topology of an amorphous\\
metal--organic framework'}
\end{center}

\tableofcontents

\clearpage
\section{Supplementary Notes}

\subsection{Supplementary Note 1: MLIP methodology}

\subsubsection{Development and composition of the reference dataset}

We generated a comprehensive dataset of ZIF structures (data locations) labelled with DFT-computed reference data (data labels), with an aim to describe disordered ZIF atomic environments accurately. The dataset development consisted of three different parts, which are described below. We visualise the overall approach and the composition of the dataset, using the style of Ref.~S\!\!\citenum{Zhou2023}, in Supplementary Figure \ref{figS0}a--b.\\

{\noindent \bf Initial `iter-0' dataset.} The aim of the initial dataset (to which we refer to `iter-0'\cite{Zhou2023}) was to begin with a broad coverage of the configurational space accessible to ZIFs, before gradually specialising towards simulations tailored to the specific needs of the application. To achieve this goal, we used three structure generation strategies. First, we `decorated and relaxed' a large set of ZIF configurations starting from {\sf AB}$_2$ framework structures, using the procedure of Ref.~S\!\!\citenum{FaureBeaulieu2023}, which we describe below. The {\sf AB}$_2$ structures were obtained from the coarse-grained {\sf AB}$_2$ MOF nets described in Ref.~S\!\!\citenum{Nicholas2021}, and the experimentally-realised zeolite framework types approved by the Structure Commission of the International Zeolite Association.\cite{Baerlocher} Second, we included DFT geometry-optimised configurations for ZIF-4, ZIF-\textbf{zni}, and ZIF-\textbf{coi} at varied (scaled) lattice parameters, including structural models with applied distortions to better sample the potential-energy surface (PES) of the relevant crystalline structures. Third, we sampled more widely the PES of the ZIF-4, ZIF-\textbf{zni}, and ZIF-\textbf{coi} crystal structures by running MD simulations with the `MOF-FF' empirical force field for ZIFs.\cite{Durholt2019} We describe these simulations below.

The decorate-and-relax procedure used, the code for which was initially introduced in Ref.~S1, was as follows. First, the connectivity between {\sf A} and {\sf B} sites was determined. Second, the structure was rescaled so that the minimum {\sf A}--{\sf B} bond length was 3.5\,\AA{}. Third, all {\sf A} sites were replaced with zinc atoms and all {\sf B} sites were replaced with imidazolate molecules. Imidazolate molecules were positioned such that the centre of the ring coincided with the {\sf B} site coordinates, and the $C_2$ symmetry axis of the molecule aligned with the vector bisecting the {\sf A}--{\sf B}--{\sf A} bond angle. Finally, the structure was optimised using the MOF-FF for ZIFs empirical force field,\cite{Durholt2019} which parameterises the system energy in terms of bonding, angular, torsional, van der Waals, and electrostatic terms. Although the force field was originally fitted for the ZIF-8 topology (\textbf{sod}), the explicit incorporation of bonding terms ensured that most hypothetical structures relaxed to chemically reasonable geometries, without overlapping atoms or excessive bond distortions. After optimising the structures, any structures with pairwise distances less than 1.0~\AA{} (or 0.6~\AA{} if including a hydrogen atom) were removed. This selection criterion ensured that highly-distorted frameworks with non-physical structures were omitted from the dataset, to prevent a skewing to structures that were less relevant.

For each configuration, three additional training structures were generated by applying random strains to the unit cells and random displacements to the atomic positions, a process colloquially referred to as `rattling':
\begin{enumerate}
	\item The cell lengths (angles) were randomly varied by up to 5~\% (2.5\textdegree) of the relaxed values, with random displacements applied to the atomic positions, having magnitudes drawn from a normal distribution with $\sigma=0.01$~\AA{}.
	\item The cell lengths (angles) were randomly varied by up to 10~\% (5.0\textdegree) of the relaxed values, with random displacements applied to the atomic positions, having magnitudes drawn from a normal distribution with $\sigma=0.1$~\AA{}.
	\item The structures were coarse-grained to {\sf AB}$_2$ nets; the cell lengths (angles) were randomly varied by up to 5~\% (2.5\textdegree), and random displacements were applied to the coarse-grained bead positions, with magnitudes drawn from a normal distribution with $\sigma=0.01$~\AA{}; and the perturbed coarse-grained structures were then decorated back to ZIF chemical units.
\end{enumerate}

Additional training structures were generated using the MOF-FF empirical potential for ZIFs.\cite{Durholt2019} These simulations aimed to broadly sample the potential energy surface rather than model specific transitions with high accuracy. First, MD simulations were conducted for the compression of ZIF-4 up to 2.0~GPa and ZIF-\textbf{zni} up to 10.0~GPa, as well as for the expansion of ZIF-\textbf{coi} under 5~GPa negative pressure. These pressures allowed for the full range of framework contraction and expansion. Training structures were extracted from these trajectories by taking evenly-spaced snapshots between the minimum and maximum volume snapshots. Second, `melted' training structures were generated by removing the bonded interactions between zinc atoms and imidazolate molecules, leaving only van der Waals and electrostatic interactions. For each of the ZIF-4, ZIF-\textbf{zni}, and ZIF-\textbf{coi} structures, simulations were performed at external pressures of 0.0 and 1.0~GPa, and at temperatures ranging from 50 to 300~K in 10~K increments. The systems were equilibrated for 10~ps, and the final trajectory snapshot was labelled using DFT and added to the training dataset. All simulations were performed with a 0.001~fs timestep in the NPT ensemble at 300~K, with pressures rapidly ramped over 5~ps.
\linebreak

{\noindent \bf Standard MD iterative training.} We sampled increasingly more disordered configurations by using iterative GAP-driven MD simulations starting from the ZIF-4, ZIF-\textbf{zni}, and ZIF-\textbf{coi} crystal structures. All simulations were performed in the NPT ensemble on single unit cells containing 272, 512, and 512 atoms for ZIF-4, ZIF-\textbf{coi}, and ZIF-\textbf{zni}, respectively. The timestep was set to 1~fs, and the temperature and pressure damping constants were set to 100~fs. Velocities were initialised randomly at 300~K. The structures were equilibrated at the initial conditions for 50~ps before heating or pressure were introduced.

For GAP-MD iterations 1--5, we performed both heating and pressurisation simulations. Heating simulations were carried out between 300 and 1,500~K at three pressure values: $-0.01$, 1, and 0.01~GPa. The temperature was ramped up over 100~ps, followed by holding at that temperature for 150~ps. Pressurisation simulations were performed between 0 and 1~GPa for ZIF-4, and 0 to 10~GPa for ZIF-\textbf{coi} and ZIF-\textbf{zni}. The pressure was ramped over 50~ps, and the structures were held for a further 200~ps.

For GAP-MD iterations 4 and 5, additional `expansion' simulations were performed for ZIF-4 and ZIF-\textbf{coi}. In these simulations, a negative pressure, ramped down from 0 to $-10$~GPa, was applied. 

In the final GAP MD iteration, `iter-6', we focused on heating ZIF-4 from 300 to 1,800~K. These temperatures were higher than those used for subsequent production runs (1,500~K), thereby providing a `buffer' to ensure that high-energy configurations were well sampled. Ten trajectories were generated with densities on a linear scale between that of crystalline ZIF-4 (tetrahedral density of 3.6~${\rm T_d~nm^{-3}}$) and ZIF-\textbf{zni} (4.7~${\rm T_d~nm^{-3}}$). The structures were heated to 1,800~K over 15~ps and equilibrated at 1,800~K for up to 250~ps. Snapshots were extracted from the equilibration period up to the point where unphysical ring-opening events occurred. Snapshots corresponding to the onset of ring-opening events were also included in the training dataset.
\linebreak

{\noindent \bf AL-HRMC active learning.} As detailed in the main text, three rounds of AL-HRMC active learning were performed. In each round, ten initial configurations were generated using the \textit{a}-\ce{SiO2} CHIK and subsequent decoration protocol outlined in the Methods section. To allow for variability in intermediate-range order within the amorphous networks, the densities of the initial random \ce{SiO2} configurations were varied linearly between 1.8 and 2.8~g~cm$^{-3}$.\cite{Erhard2022} This approach produced ZIF structures with distinct ring statistics and corresponding pore shapes and sizes. For each input configuration, three HRMC refinements were performed with $\sigma \in \{0.1,0.01,0.001\}$ to adjust the relative weighting of the data-fitting terms compared to the energy. Smaller values of $\sigma$ increased the weighting of the data-fitting component, encouraging the exploration of more extrapolative atomic environments.
	
In theory, an HRMC simulation could be terminated as soon as an extrapolative environment was identified, allowing the final snapshot to be extracted, labeled, and used for retraining the ML potential. In practice, we allowed the HRMC simulations to progress for 1 million steps, and stored snapshots from the simulation every 10,000 proposed moves. During a post-processing stage, for each configuration in the stored trajectory, we summed the $\gamma_{\rm max}$ values for each species to assign a per-structure $\gamma$ value. To select configurations for labelling, we performed automated clustering on the $\gamma$ values across the trajectory using the \texttt{DBSCAN} algorithm (\texttt{eps} = 0.1, \texttt{min\_samples} = 2). The final configuration was then selected from each cluster.
\linebreak

{\noindent \bf Overall dataset.}
The composition of the final dataset is visualised in Supplementary Figure \ref{figS0}b. In this two-dimensional structure map, the distance between any two points indicates the dissimilarity between those configurations, as determined by the ACE $B$-basis vectors, averaged over all zinc environments in each structure. Structures that are close in the map are similar; structures that are far apart are dissimilar.

The dataset is designed to encompass a wide range of configurations. Regions labelled (1) and (2) predominantly represent crystalline structures similar to ZIF-4 (\textbf{cag}) and ZIF-\textbf{zni}, respectively. These regions were particularly explored in the early stages of MD iterative training, as indicated by the high density of blue markers. Many crystalline hypothetical ZIF ($h$-ZIF) structures are also located in these regions. Region (3) includes dense frameworks, mostly explored during high-pressure iterative MD training. Region (4) also contains dense frameworks generated during iter-0 but is distinct from region (3) due to highly distorted local environments, resulting from decorating networks with topologies that do not readily support the Zn(Im)$_2$ structural motif. Region (5) contains `liquid'-like structures with many under-coordinated zinc sites. Region (6) hosts $a$-ZIF snapshots, with most configurations generated during AL-HRMC. Interestingly, this disordered region of configuration space bridges regions (1) and (2), mirroring the transformation of ZIF-4 to ZIF-\textbf{zni} \textit{via} $a$-ZIF.

We visualise the diversity in configurations by colour-coding the map using structural and energetic information in Supplementary Figure \ref{figS0}c--g, extending upon ideas in Ref.~S\!\!\citenum{Nicholas2020}. The dataset spans a wide range of tetrahedral densities, encompassing all known experimental Zn(Im)$_2$ values\cite{Zheng2023} (panel c). The highest-energy structures generally feature defects such as under-coordinated sites, as found in the liquid-like configurations in regions (4) and (5) (panels d and e). ZIF-4 and ZIF-\textbf{zni} are important structural markers in the ZIF configuration space. The dissimilarity of all structures with ZIF-4 and ZIF-\textbf{zni} is shown in panels f and g, respectively. The dissimilarity is defined as the Euclidean distance between the averaged zinc ACE vectors of each structure and the reference structure. Smaller distances (lighter shades) indicate more similar structures, and vice versa. Many disordered configurations, including representative $a$-ZIF snapshots, are found at intermediate dissimilarities between these two crystalline phases.
\linebreak

{\noindent \bf Testing set.} The testing set was generated by performing MD simulations using the ACE model fitted after the final round of HRMC-iterative training. For each crystalline Zn(Im)$_2$ polymorph with topologies \textbf{cag}, \textbf{coi}, \textbf{crb}, \textbf{dft}, \textbf{gis}, \textbf{mer}, \textbf{neb}, \textbf{nog}, \textbf{sod}, \textbf{zni}, 10 starting configurations were created by rescaling the densities linearly between the experimental density of each polymorph and 1.63~g~cm$^{-3}$, which is the density reported for ZIF glasses.\cite{Gaillac2017} This approach samples the full range of expected densities.
	
Each structure was then subjected to a melting simulation. The simulation protocol consisted of MD simulations in the NVT ensemble, divided into two stages: an equilibration stage and a melting stage. For the equilibration stage, each structure was held at 20~K for 5~ps with a temperature damping constant of 0.01~ps and a time step of 0.25~fs. The temperature was then ramped from 20~K to 300~K over 224~ps with a temperature damping constant of 0.1~ps. The structure was subsequently held at 300~K for 5~ps. For the melting stage, the temperature was incrementally increased from 300~K to 1,500~K in 12 stages, each involving 200~ps of ramping followed by 200~ps of equilibration, using a time step of 0.1~fs and a temperature damping constant of 0.04~ps. At the end of each stage, the configuration was extracted, labelled with DFT energies and forces, and added to the testing set.
	
To include topologically disordered structures, ten \textit{a}-\ce{SiO2} configurations containing 64 formula units were generated following the CHIK melt-quench protocol described in the Methods section, and decorated to form ZIF structures. The disordered ZIF structures were then subjected to the same equilibration and melting simulations described above. However, because these structures contain 1,088 atoms each, DFT calculations are considerably more computationally expensive than those for the crystalline ZIF structures (which typically contain between 272 and 544 atoms per unit cell). As a result, only the configurations equilibrated at 300~K and 1,500~K were labelled with DFT and added to the testing set.

\clearpage
\subsubsection{GAP model fitting}
The initial versions of the potential were developed using the Gaussian Approximation Potential (GAP) framework.\cite{Bartok2010,Bartok2013,Deringer2021} In this discussion, we describe the GAP model hyperparameters which were determined by fitting different potentials to the labelled iter-0 dataset, and selecting the values of the hyperparameters based to the resulting root-mean-square error (RMSE) of the energy per atom. A summary of the final settings is shown in Supplementary Table 2.

We incrementally added 2-body, 3-body, and many-body SOAP descriptors to our model, setting the weight, $\delta$, for each descriptor as the RMSE of the energy per atom of the model before incorporating that specific body-order descriptor.\cite{Deringer2017, Deringer2021} In this way, the $\delta$ are scaling parameters, and each corresponds to the distribution of energy contributions that a given interaction term (descriptor) has to represent. 

Initially, we fitted a two-body-only model with all radial cut-offs, $r_{\rm cut}$ set to 2.4\,\AA{}, which included the first and second nearest neighbour atoms. This cut-off was later increased to 6.0\,\AA{} for GAP-iter-4 during iterative refinement, as this adjustment improved the numerical accuracy of the model. Next, we added intramolecular three-body descriptors: \{H, C, N\}, \{C, N, N\}, \{H, C, C\}, and \{C, C, N\}, with $r_{\rm cut}=1.8$\,\AA{}. Each set of curly braces denotes a unique descriptor, including all triplets of the specified species within the cut-off distance. Additionally, we added intermolecular three-body descriptors with $r_{\rm cut}=2.2$\,\AA{}: \{C, N, Zn\} and \{N, N, Zn\}. 

Finally, we added a many-body SOAP descriptor for each chemical species.\cite{Bartok2013} We used the \texttt{TurboSOAP} implementation of SOAP,\cite{Caro2019} which enabled the optimisation of the per-species smoothing parameters, $\sigma_{\rm at}$. The values of each $\sigma_{{\rm at}, \mu}$ were determined for each species $\mu$ by varying each parameter in turn on a linear scale between 0.1 and 1.0~\AA{} with a step size of 0.05~\AA{}, keeping all other $\sigma_{{\rm at}, \mu}$ fixed, and selecting the value that resulted in the smallest energy training error.

During the GAP model fitting procedure, rather than using all $N_{\rm total}$ atomic environments in the dataset, $N_{t}$ representative training points are selected for each descriptor to describe the dataset. This process is referred to as `sparsification' (see Ref.~S\!\!\citenum{Deringer2021} and references therein). We used different numbers of representative points from various parts of the dataset to reflect the diversity of atomic environments in each section and the relative importance of those configurations for the application domain. For sparsification of the two- and three-body terms, a uniform grid of $N_{\rm sparse}$ basis function locations was used.\cite{Bartok2010} For the SOAP kernel, sparsification was achieved using CUR decomposition, a matrix factorisation technique that selects a subset of representative columns and rows.\cite{Mahoney2009} The number of sparse points for each descriptor is given in Supplementary Table 2.

In the context of GAP models, regularisation helps to control the `expected error' of the input data by applying parameters that penalise overfitting, thus ensuring the model generalises well to unseen data. The regularisation parameters $\sigma_{\rm energy}$, $\sigma_{\rm force}$, and $\sigma_{\rm stress}$ represent the expected uncertainty in the energy, force, and stress predictions, respectively. Lower values indicate higher confidence in the accuracy of the input data, while higher values suggest greater uncertainty and flexibility in fitting. We divided our dataset into categories to reflect these differences in accuracy, as shown in Supplementary Table 3. For example, crystalline structures, which have well-defined local environments, were assigned small regularisation values, reflecting higher confidence in the data. In contrast, more diverse categories like `liquid-like' structures and `buffer' configurations that feature structure degradation events were assigned higher regularisation values, indicating a higher tolerance for variation and uncertainty in these data types.\\

{\noindent \bf Computational details.} For fitting in practice we used the MPI-parallelised version of {\tt gap\_fit}.\cite{Klawohn2023} This enabled the distribution of a total memory requirement of 15~TB across 32 nodes (4,096 CPU cores) of the ARCHER2 UK national supercomputing facility.\cite{Archer2}

\clearpage
\subsubsection{ACE model fitting}
The final ML interatomic potential models were trained using the ACE framework.\cite{Drautz2019,Lysogorskiy2021,Bochkarev2022} ACE decomposes the system energy into body-ordered contributions, which are then summed into local atomic energies. It does this by first projecting the atomic neighbour density onto one-particle basis functions, $\phi_{\boldsymbol{\nu}}$. The product of these projections containing $\boldsymbol{\nu}$ factors gives a ($\nu$+1)-body `$A$-basis' function. By averaging the $A$-basis function over the three-dimensional rotation group, a rotationally-invariant `$B$-basis' is formed. An atomic property (index $p$) of atom $i$ is then defined as
\begin{equation}\label{eq:ace-basis-set}
	\varphi_i^{(p)} = \sum_{\boldsymbol{\nu}} c_{\boldsymbol{\nu}}^{(p)} \boldsymbol{B}_{i\boldsymbol{\nu}},
\end{equation}
where $c_{\boldsymbol{\nu}}^{(p)}$ are the expansion coefficients which share the multi-indices $\nu$ with $\boldsymbol{B}_{i\boldsymbol{\nu}}$, which describe the list of basis functions in the cluster. The atomic energy can be expressed as a function of $P$ such atomic properties:
\begin{equation}\label{eq:ace-atomic-properties}
	\epsilon_i = \mathcal{F}\left( \varphi_i^{(1)}, ..., \varphi_i^{(P)} \right),
\end{equation}
where $\mathcal{F}$ is a general non-linear embedding function. For example, the Finnis--Sinclair potential \cite{Finnis1984} is expressed in the form:
\begin{equation}\label{eq:fs}
	\epsilon_i = \varphi_i^{(1)} + \sqrt{\varphi_i^{(2)}}.
\end{equation}
Here, the linear term, $\varphi_i^{(1)}$, represents the repulsive core-core interaction, and the square-root term, $\sqrt{\varphi_i^{(2)}}$, represents the cohesive, many-body nature of metallic bonding.  Incorporating additional atomic properties introduces flexibility into the functional form.\cite{Bochkarev2022,Erhard2024}

The loss function used in the fitting software we use, \texttt{pacemaker}, \cite{Lysogorskiy2021, Bochkarev2022} consists a weighted mean square error of the difference between energies and forces in the reference dataset and those predicted by ACE:
\begin{equation}
	\mathcal{L} = (1 - \kappa)  {\rm \Delta}_{E}^2 + \kappa {\rm \Delta}_{F}^2 + {\rm \Delta}_{\rm coeff} + {\rm \Delta}_{\rm rad},
\end{equation}
where the relative contribution of the errors of the energies ${\rm \Delta}_{E}^2$ and forces ${\rm \Delta}_{F}^2$ is weighted by the parameter $\kappa$. For all ACE models fitted in the present work, we performed four consecutive fits (an approach referred to as `upfitting'\cite{ThomasduToit2024}) using $\kappa=0.95$, 0.1, 0.01, and 0.001, thereby gradually increasing the relative weighting of the energies in the loss function. ${\rm \Delta}_{\rm coeff}$ and ${\rm \Delta}_{\rm rad}$ are regularisation contributions which ensure the smoothness of the PES and the radial basis functions, respectively. These were both set to a (default) value of $10^{-8}$.\\

{\noindent \bf Initial ACE hyperparameter values.} For the initial ACE models fitted and used during the ACE-driven MD and AL-HRMC iterations, we kept the hyperparameters fixed. For all species, the atomic neighbour density was determined up to a cut-off of $r_{\rm cut}=6.0$~\AA{}, to maintain consistency with the GAP models, with the density smoothly tapered to zero at the cut-off over a distance of \texttt{dcut}~$=0.01$~\AA{}. A Finnis--Sinclair embedding was used ($P = 2$) with 1,200 basis functions per element (4,800 total functions).\\

{\noindent \bf ACE extrapolation grade.} In ACE, the `extrapolation grade', $\gamma$, determines whether a given atomic environment can be regarded as an interpolation or extrapolation of the reference training dataset.\cite{Lysogorskiy2023} The metric is based on the D-optimality criterion.\cite{Podryabinkin2017} The method constructs an `active set' matrix, $\hat{A}_{\mu}$, for each chemical species, $\mu$, using the MaxVol algorithm, which contains a subset of the ACE basis vectors computed for all atoms in the training dataset which span the largest volume. The extrapolation grade for atom $i$ of type $\mu$ is then defined as
\begin{equation}
	\gamma_i = \max_k{(|\gamma_k^{(i)}|)} = \max \left| \boldsymbol{B}_i \cdot \hat{A}_{\mu}^{-1} \right|.
\end{equation}
The \texttt{pacemaker} software offers both linear and non-linear implementations of the D-optimality criterion. In the present work, we used the full non-linear embedding.\\

{\noindent \bf ACE model hyperparameter optimisation.} After finalising the training dataset, we performed hyperparameter optimisation for the ACE models, automated using the XPOT code\cite{ThomasduToit2023,ThomasduToit2024} which interfaces to \texttt{scikit-optimize} for the Bayesian optimisation routines.

In the present work, we optimised ACE hyperparameters for models with different $P \in \{2,4,6\}$ atomic properties and number of basis functions per element to explore different trade-offs between numerical accuracy and model efficiency.\cite{Erhard2024,ThomasduToit2024} In total, we created 6 models which are described in Supplementary Table 4. For all models, the functional form of the first two atomic properties were set to the Finnis--Sinclair form (Eq.~\ref{eq:fs}). For models with $P>2$, the exponents of the additional atomic properties ($f_{i>2}$) were optimised. We also optimised the hyperparameters \texttt{dcut} (which controls the range over which the radial components smoothly transition from one to zero) and \texttt{wlow} (which adjusts the relative weighting of low-to-high energy structures in the dataset). Additionally, the choice of radial basis function and associated parameters were optimised, with all models ultimately refining to use simplified spherical Bessel functions. The final optimised parameters for each model are given in Supplementary Table 4.

\clearpage
\subsubsection{Validation for crystalline polymorphs}
\label{sec:crystal-validation}
While we have focused on applying the AL-HRMC approach to find sensible amorphous structures, the training dataset includes the crystalline Zn(Im)$_2$ phases. In this section, we show the MLIP models are capable of accurately reproducing the DFT reference data for these crystalline materials.\\

{\noindent \bf Energy-volume curves.} Supplementary Figure \ref{figS5}a shows the energy--volume (EV) curves of ZIF-4 (\textit{top}), ZIF-\textbf{coi} (\textit{middle}), and ZIF-\textbf{zni} (\textit{bottom}) computed with DFT (black circles) at the SCAN meta-GGA level of theory. The EV curves were obtained by rescaling the DFT-minimised reference structures. For each configuration, a fixed-volume, variable-cell relaxation (VASP \texttt{ISIF=4}) was performed without symmetry constraints. This was followed by a second relaxation, during which the cell was held fixed at the previously relaxed shape while the atomic positions were optimised (VASP setting \texttt{ISIF=2}). In both stages of geometry optimisation, the Kohn--Sham equations were solved self-consistently until both the total free energy change and the band-structure-energy change converged to within $10^{-8}$~eV per cell. A plane wave cut-off energy of 1,000~eV was used, with partial occupancies treated using Gaussian smearing ($\sigma=0.2$~eV). The force criterion for a converged geometry was set such that the maximum force on any atom was below 0.01~eV/\AA{}. Final energies were obtained \textit{via} a subsequent single-point calculation. 

Notably, the energy difference between the ZIF‐\textbf{zni} and ZIF‐\textbf{coi} polymorphs is subtle,\cite{Schroder2013} and given that these small differences depend sensitively on the choice of functional and dispersion corrections, we do not assign strong weight to the absolute quantitative analysis. Instead, our focus is on assessing how well the ML model has learned from the training data, as indicated by its ability to reproduce the DFT‐derived EV curves.

We found that the ACE models' ability to reproduce the energies for the configurations relevant to the EV curves could be systematically improved by increasing the weighting of crystalline configurations within the dataset. To achieve this, we up-fitted the final ACE model 3 discussed earlier, keeping all hyperparameters fixed (including $\kappa=0.001$). The energy and force weightings were set uniformly across the dataset, except for the crystalline configurations, which were systematically up-weighted by factors ranging from 5 to 100 times relative to the rest of the dataset. We allowed the fit to continue for 1,000 training steps.

The systematic improvement is shown in Supplementary Figure \ref{figS5}a, where progressively darker lines correspond to higher weightings of the crystalline configurations, and better reproduce the reference DFT energies. We further quantify this improvement in terms of the energy RMSE in Supplementary Figure \ref{figS5}b. Additionally, we confirmed that the upfitting process did not significantly affect the energy predictions for the $a$-ZIF models. This was verified by calculating the energy RMSE using the predictions from ACE model 3 as the reference values, with the corresponding results displayed in grey in Supplementary Figure \ref{figS5}b. In particular, the model with 100$\times$ upweighting effectively captures the subtle difference in EV curves for the \textbf{zni} and \textbf{coi} polymorphs in response to density variations (Supplementary Figure \ref{figS5}c), as described by the reference DFT method.\\

To further evaluate the behaviour of our ML potential model, we performed MD simulations in the NPT ensemble using \texttt{LAMMPS}\cite{Thompson2022} to monitor the evolution of lattice parameters for the ZIF‐\textbf{zni} system under pressures up to 0.75~GPa. A Nosé–Hoover thermostat and barostat\cite{Nose1984,Hoover1985} were used with temperature and pressure damping parameters of 100~fs and 1,000~fs, respectively, and a timestep of 0.25~fs. Simulations were initialised with single unit cells. We find a good agreement with the experimental lattice parameters reported in Ref.~S\!\!\citenum{Spencer2009}, as shown in Supplementary Figure \ref{figS5}d. We note that we deliberately did not simulate at larger pressures, as a single-crystal-to-single-crystal phase transition is known to occur,\cite{Spencer2009} introducing challenges---such as the need for enhanced sampling techniques---that are beyond the scope of this work.

\clearpage
\subsection{Supplementary Note 2: Effect of annealing temperature}
\label{sec:annealing-temperature}
The final AL-HRMC $a$-ZIF models were obtained through a two-step process: (1) a high-temperature melt-quench MD simulation starting from a disordered network (see Methods) and (2) HRMC refinements to generate the final $a$-ZIF structures. In the main text, we reported results using a 1,500~K annealing temperature. To verify that our results are not strongly dependent on the annealing temperature, we performed additional melt-quench and HRMC simulations, with 1,200, 1,300, and 1,400~K annealing temperatures.

The fits to experimental X-ray and neutron total scattering functions\cite{Bennett2010} for these simulations are presented in Supplementary Figure~\ref{figS3}a and b, respectively. The results demonstrate that the models reproduce the experimental total scattering functions consistently, regardless of the annealing temperature.

The purpose of the MD melt-quench simulation preceding HRMC refinements was to facilitate topological rearrangement. As shown in Supplementary Figure~\ref{figS3}c, the choice of annealing temperature does not significantly affect the amorphous topology metrics for $a$-ZIF.

\clearpage
\subsection{Supplementary Note 3: Comparing RMC and AL-HRMC \textit{a}-ZIF models}
\label{sec:compare-rmc-zif-ace}
In this section, we provide a comparison between the \textit{a}-ZIF structural model produced by a `conventional' RMC refinement\cite{Bennett2010} and the model reported in the present work.\\

{\noindent \bf Fit to experimental data.} The final fits to the experimental X-ray and neutron total scattering functions from Ref.~S\!\!\citenum{Bennett2010} are shown in Supplementary Figure \ref{figS4}a and b, respectively. Both functions are better reproduced by the AL-HRMC model than the RMC model, which we quantify using the $\chi^2$ metric in Supplementary Table \ref{tab:chi2_comparison}.\\

{\noindent \bf Bond-length and bond-angle distributions.} In panels (c) and (d) of Supplementary Figure \ref{figS3}, we present the key bond-length and bond-angle distributions for the two models. Both models exhibit distributions centered around characteristic lengths and angles typical of a tetrahedral network, as expected. However, notable differences emerge in the details.

The Zn--N bond length in the ZIF-ACE model is slightly shifted towards smaller $r$ values compared to the RMC model, with a broader distribution. The Zn--Zn correlation is particularly striking, showing a significantly broader distribution in the ZIF-ACE model, with a pronounced tail extending towards smaller $r$. This likely reflects the lack of a fixed-topology constraint in the ZIF-ACE model, as opposed to the RMC model, where distance windows were used to enforce network topology. This interpretation aligns with the observation of a small fraction of under-coordinated sites in the ZIF-ACE model.

The bond-angle distributions for N--Zn--N and Zn--Im--Zn show good agreement between the two models. However, there is a marked difference in the Im--Zn--Im bond-angle distributions. In the ZIF-ACE model, the Im--Zn--Im distribution closely matches the N--Zn--N distribution, consistent with an imidazolate ligand bridging two zinc atoms with minimal out-of-plane rotation, which is energetically unfavourable. In contrast, the RMC model displays significant deviations from this expected bonding arrangement, the formation of which can likely be attributed to the lack of energetic constraints in the refinement process. These differences highlight the particular strength of energetic considerations when modelling molecular systems. Such considerations not only ensure physically reasonable structures but also provide essential insights into higher-order correlations that cannot be inferred from diffraction experiments alone.\\

{\noindent \bf Local energy distributions.} We show in Supplementary Figure \ref{figS4}e the per-species local energy distributions, computed using the ZIF-ACE-24 potential (Supplementary Table \ref{tab:ace-xpot}, model 3), for both the RMC and ZIF-ACE models. Overall, the energy distributions for the ZIF-ACE models are slightly shifted towards lower energies compared to the RMC models and exhibit narrower distributions, indicating a more energetically favourable structure.

A notable exception is observed in the zinc distribution, where the ZIF-ACE model exhibits a broader energy spread than the RMC model. This broader distribution arises due to the presence of over-coordinated (five-fold) and under-coordinated (three-fold) zinc sites in the ZIF-ACE model, in contrast to the RMC model, which exclusively features four-fold coordinated zinc sites. To further illustrate this, the zinc local energy distributions, categorised by coordination number, are shown in Supplementary Figure \ref{figS4}f.

Although we do not assign specific physical significance to the local energy values, these distributions provide insight into the structural differences between the models, particularly the broader range of coordination environments captured in the ZIF-ACE model. This variation explains the broader zinc energy distribution observed in the ZIF-ACE model compared to the more constrained RMC model.\\

\clearpage
\subsection{Supplementary Note 4: Cumulative all-rings vertex symbol}

{\noindent \bf Definitions.} For each node in a network, we compute a Cumulative All-Rings Vertex Symbol (CARVS), denoted $\bf \bar{c}$, as follows:
\begin{enumerate}

    \item {\bf Ring enumeration.} For a given node $j$, enumerate all rings including $j$ and assemble them into a vector ${\bf c}^{(j)}: {\bf c}^{(j)} = \left[ a_1^{(j)}, a_2^{(j)}, \dots, a_S^{(j)} \right]$ where each element $a_i^{(j)}$ represents the number of rings of size $i$ around node $j$, up to a maximum ring size $S$.

    \item {\bf Matrix representation.} For a structure consisting of $N$ nodes, we define a matrix $\bf C$ of size $N \times S$, where the element $C_{ji}$ indicates the number of rings of size $i$ around node $j$.

    \item {\bf CARVS vector.} The CARVS vector $\bf \bar{c}$ is computed as the column-wise mean of $\bf C$:

        \begin{equation}
            \bar{c_i} = \frac{1}{N} \sum_{j=1}^N C_{ji}, \quad  \text{ for } i = 1,~2,~\dots,~S.
        \end{equation}
        
\end{enumerate}

To compute the standard deviation among node environments, we first calculate the Euclidean distance between each node's ${\bf c}^{(j)}$ and the average vector $\bf \bar{c}$:
\begin{equation}
    d^{(j)} = \left| {\bf c}^{(j)} - {\bf \bar{c}} \right|.
\end{equation}
The standard deviation $\sigma$ is then calculated as
\begin{equation}
    \sigma = \sqrt{ \frac{1}{N} \sum_{j=1}^{N} \left( d^{(j)} \right)^2 }.
\end{equation}

The CARVS symbol can be used to define a distance metric that quantifies the topological dissimilarity between two structures $\alpha$ and $\beta$.\\

{\noindent \bf Relationship between CARVS and total ring counts.} The CARVS represents the average number of rings of a given size, $s$, per node in a network. It is therefore related to the total number of rings of size $s$, denoted $R_s$, by a simple scale factor:
\begin{equation}
    R_s = \left(\frac{N}{s}\right) {\bf \bar{c}}_s
\end{equation}
where multiplication by the total number of nodes $N$ gives the total fractional contributions across all nodes to the ring count, and the division by a factor of $s$ reflects the fact that $s$ nodes contribute to a ring of size $s$.\\

{\noindent \bf Relationship between CARVS and vertex symbols.} Here, we compare the CARVS and vertex symbols for a selection of \textit{crystalline} ZIF polymorphs. 

For an $n$-coordinated vertex in a three-dimensional net, there are $n(n-1)/2$ angles. Crystalline ZIF nets feature 4-connected nodes, meaning they possess $4(4-1)/2=6$ angles. The vertex symbol takes the form $[A_a.B_b.C_c. \cdots]$, where $a$ denotes the number of shortest rings that are $A$-membered at the first angle, $b$ represents the number of shortest rings that are $B$-membered at the next angle, and so on.\cite{Blatov2010} If a given angle is associated with a single shortest ring, the subscript is omitted. In the case of 4-coordinated nets, the angles are grouped into three pairs of opposite angles.\cite{OKeeffe1991} The vertex symbol is then arranged in lexicographic order, respecting this pairing.

We present the vertex symbols for the \textbf{cag}, \textbf{zni}, \textbf{sod}, \textbf{zec}, \textbf{coi}, and \textbf{nog} nets in Supplementary Table \ref{tab:topology-symbols}. The \textbf{cag}, \textbf{zni}, and \textbf{sod} nets are uninodal, meaning each possesses a single vertex symbol. The \textbf{zec} net, in contrast, is binodal with a stoichiometry of 1:4, while \textbf{coi} is 4-nodal with a stoichiometry of 1:1:1:1 and \textbf{nog} is a 5-nodal net with a stoichiometry of 1:1:1:1:1. For clarity, the multiplicity of each node is indicated as a prefix to the vertex symbol in Supplementary Table \ref{tab:topology-symbols}.

It is also possible to extend the vertex symbol to describe all rings at each angle, rather than just the shortest one. These extended vertex symbols, which more closely resemble the CARVS, are also provided in Supplementary Table \ref{tab:topology-symbols}. By enumerating all angles and ordering the resulting symbol lexicographically, the per-node CARVS vector, ${\bf c}^{(j)}$, can be determined. In multi-nodal nets, the CARVS is computed as the weighted average of ${\bf c}^{(j)}$ over all unique nodes, where each node's contribution is weighted by its multiplicity. For uninodal nets, the CARVS simplifies to ${\bf c}^{(j)} = {\bf \bar{c}}$, yielding a standard deviation of $\sigma = 0$.

\clearpage
\subsection{Supplementary Note 5: Ring-size energetics in tetrahedral networks}

Our \textit{a}-ZIF model exhibits a broad ring-size distribution, spanning from three to approximately 15 nodes. Since no crystalline ZIF polymorphs feature 3-membered rings, we examined whether such topological features are expected in the amorphous phase by analysing local node energies as a function of ring size.\\

{\noindent \bf Model accuracy.}  
To validate the accuracy of our MLIP model in describing the range of topological motifs present in \textit{a}-ZIF, we evaluated the force RMSE of zinc nodes as a function of ring size using the disordered structures from our test set (see Supplementary Note 1). Supplementary Figure \ref{figS7}a confirms that the model reliably describes node environments across different ring sizes, including 3-membered rings.\\

{\noindent \bf Relative ring energies.}  
Supplementary Figure \ref{figS7}b presents the average node energy as a function of ring size for the topologically disordered test set. While 3-membered rings exhibit a slight energetic penalty, the relative energy scale remains narrow, suggesting a low barrier to their formation.

For larger structural models, we show the average node energy as a function of ring size in \textit{a}-Si, \textit{a}-\ce{SiO2}, and \textit{a}-ZIF (Supplementary Figure \ref{figS8}a–c). These calculations, performed using the MLIP model employed for structural generation, reveal similar trends across materials. Despite the energetic penalty, small (3-membered) rings are observed in all cases. Notably, rare `2-membered' rings—where two nodes are bridged by two linkers—appear in both \textit{a}-\ce{SiO2} and \textit{a}-ZIF. These configurations are highly energetic and likely arise as artefacts of the high-temperature annealing process used to generate the models.

While local energetics are strongly influenced by coordination number (see Supplementary Note 5), the test set structures contain only 4-fold connected zinc nodes. The similarity between their energy distributions (Supplementary Figure \ref{figS7}b) and those of larger \textit{a}-ZIF models (Supplementary Figure \ref{figS8}a), which include defects, suggests that these trends are largely independent of coordination number.

\clearpage
\section{Supplementary Figures}

\begin{figure}
    \centering
    \includegraphics{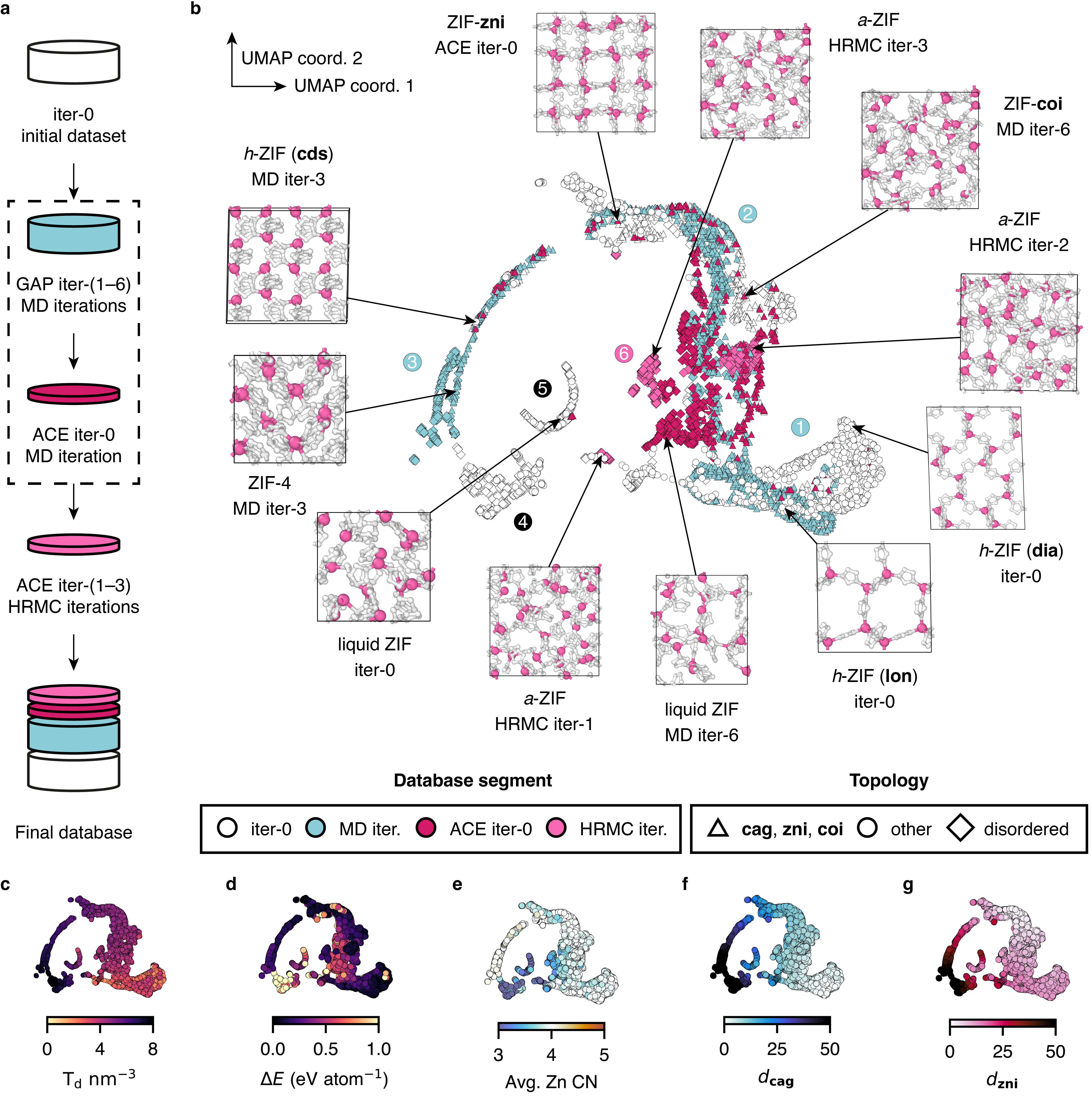}
    \caption{
        \textbf{A diverse ZIF dataset for fitting MLIPs.}
        (\textbf{a}) The three components of our ZIF ML potential dataset: \texttt{iter-0} structures, MD-derived configurations, and AL-HRMC configurations. (\textbf{b}) A two-dimensional structure map based on zinc-centred ACE vectors embedded using UMAP,\cite{Mcinnes2018} coloured by dataset segment. Triangle markers indicate key crystalline topologies (\textbf{cag}, \textbf{zni}, \textbf{coi}), circles denote other detectable topologies,\cite{Zoubritzky2022} and diamonds represent disordered structures. (\textbf{c}--\textbf{g}) The same map colour-coded by framework density, relative energy, average Zn coordination number (CN), and dissimilarity to ZIF-4 (\textbf{cag} topology) as well as ZIF-\textbf{zni} (based on ACE vector distance), respectively. Structure images were created using OVITO,\cite{Stukowski2010} with hydrogen atoms omitted for clarity.
    }
    \label{figS0}
\end{figure}

\begin{figure}
    \centering
    \includegraphics[width=\linewidth]{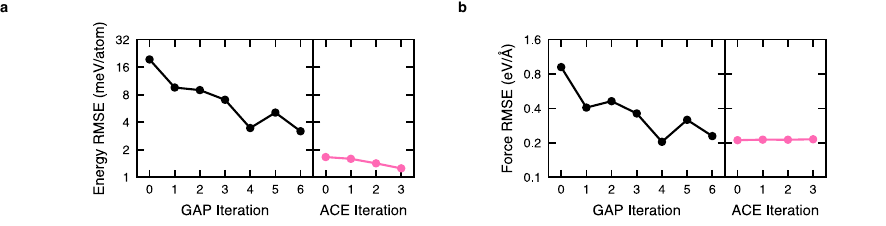}
    \caption{
    \textbf{Numerical validation during iterative training.}
    The (\textbf{a}) energy RMSE and (\textbf{b}) force RMSE during the GAP-based MD iterative training (black) and AL-HRMC iterations (pink) evaluated on the test set. 
    }
    \label{figS1}
\end{figure}

\begin{figure}
    \centering
    \includegraphics[width=\linewidth]{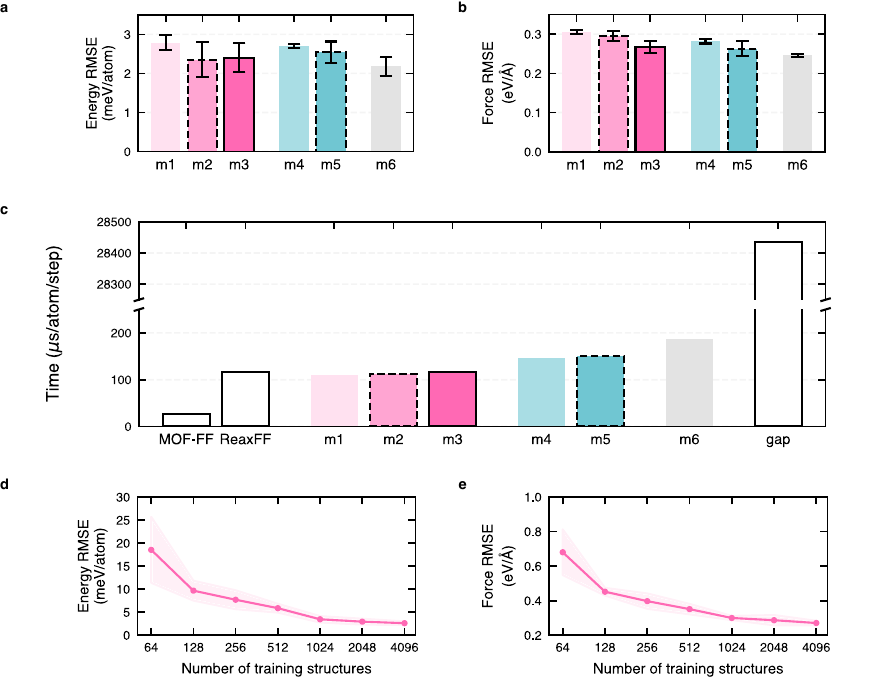}
    \caption{
    \textbf{Hyperparameter optimisation and numerical validation for ZIF ACE models.}
    The (\textbf{a}) energy RMSE and (\textbf{b}) force RMSE for the optimised ACE models with different numbers of atomic properties and basis functions (see Supplementary Table \ref{tab:ace-xpot}). For each model, 5 independent fits were performed with different random seeds. The bar heights indicate the mean RMSE, and the whiskers indicate the standard deviation.
    (\textbf{c}) Comparing the (CPU) evaluation time of each ACE model during MD simulations to the GAP model from this work. We also include the MOF-FF for ZIFs non-reactive empirical force field\cite{Durholt2019} and the reactive ReaxFF force field\cite{Yang2018} for reference.
    Having selected m3 as a good compromise between accuracy and computational efficiency, we confirmed our model was saturated with respect to training data by fitting models with progressively fewer training configurations, and evaluated the (\textbf{d}) energy and (\textbf{e}) force RMSE. 
    }
    \label{figS2}
\end{figure}

\begin{figure}
    \centering
    \includegraphics[width=\linewidth]{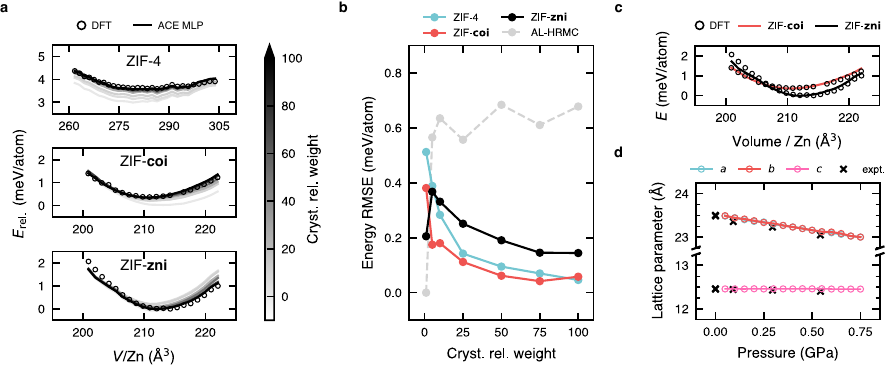}
    \caption{
    \textbf{Up-fitting the ACE ML potential model for crystalline structures.}
    (\textbf{a}) By systematically up-weighting the crystalline ZIF-4, ZIF-\textbf{coi}, and ZIF-\textbf{zni} structures in a subsequent up-fitting of the ACE model 3, the models (lines) better reproduce the reference DFT energy--volume curves (black circles). Darker shading corresponds to models with larger weights applied to the crystalline configurations during ML potential training.
    (\textbf{b}) The difference between the ACE ML potential model predictions and DFT reference values as a function of different relative crystalline weights, for ZIF-4, ZIF-\textbf{coi}, and ZIF-\textbf{zni}. All other data locations were weighted evenly. We confirm that the up-weighting of the crystalline structures does not significantly worsen the prediction for the amorphous ZIF models by showing the RMSE energy between the ACE model 3 and the subsequent up-fitted models. In all cases, the predicted energy is $<$ 1~meV/atom.
    (\textbf{c}) The up-weighted model better captures the subtle energy landscape between the dense ZIF-\textbf{coi}, and ZIF-\textbf{zni} crystalline polymorphs.\cite{Spencer2009,Schroder2013}
    (\textbf{d}) Evolution of the ZIF-\textbf{zni} lattice parameters under increasing pressure in MD simulations, compared to experimental data reported in ref.~S\!\!\citenum{Spencer2009}.
    }
    \label{figS5}
\end{figure}

\begin{figure}
    \centering
    \includegraphics[width=\linewidth]{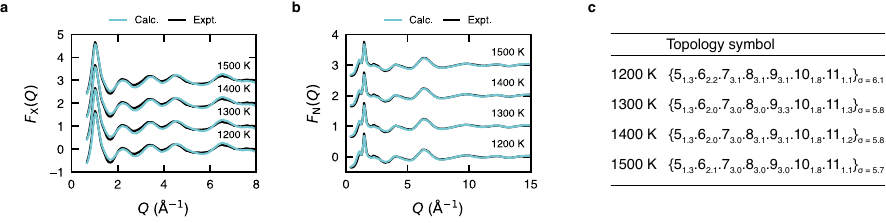}
    \caption{
    \textbf{Effect of MD annealing temperature on AL-HRMC models.}
    The fits to the (\textbf{a}) X-ray and (\textbf{b}) neutron total scattering data from Ref.~S\!\!\citenum{Bennett2010} were of comparable quality when the MD annealing was conducted at 1,200, 1,300, and 1,400~K, as compared to the fits reported in the main text, which were performed at 1,500~K.
    (\textbf{c}) The topology metrics (see main text) were also consistent for the different annealing temperatures.
    }
    \label{figS3}
\end{figure}

\begin{figure}
    \centering
    \includegraphics[width=\linewidth]{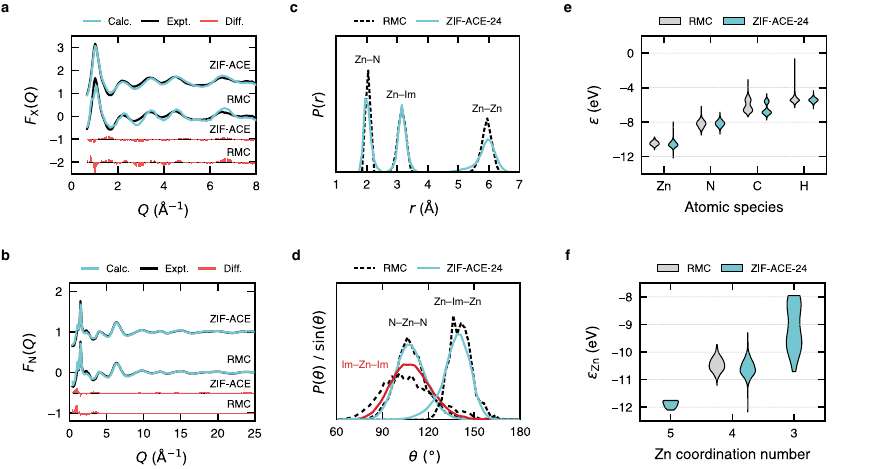}
    \caption{
    \textbf{Comparison of} {\bf\textit{a}}\textbf{-ZIF models from this work and an RMC model.}
    (\textbf{a}) X-ray and (\textbf{b}) neutron structure factors from experiments (black),\cite{Bennett2010} our \textit{a}-ZIF models, and the RMC model of Ref.~S\!\!\citenum{Bennett2010}. The difference between each model and the experiment are shown in red.
    (\textbf{c}) Bond-length and (\textbf{d}) bond-angle distributions for the RMC (dashed black line) and our \textit{a}-ZIF models (blue solid line).
    (\textbf{e}) Local energy distributions, separated by atomic species, for the RMC model compared to our \textit{a}-ZIF models.
    (\textbf{f}) Comparing the zinc local energies for our \textit{a}-ZIF models and the RMC model, separated by the number of bound nitrogen atoms (a cutoff of $r_{\rm cut}=2.5$~\AA{} was used to determine Zn--N neighbours). We note the RMC model only had four-fold connected zinc sites.
    }
    \label{figS4}
\end{figure}

\begin{figure}
    \centering
    \includegraphics[width=\linewidth]{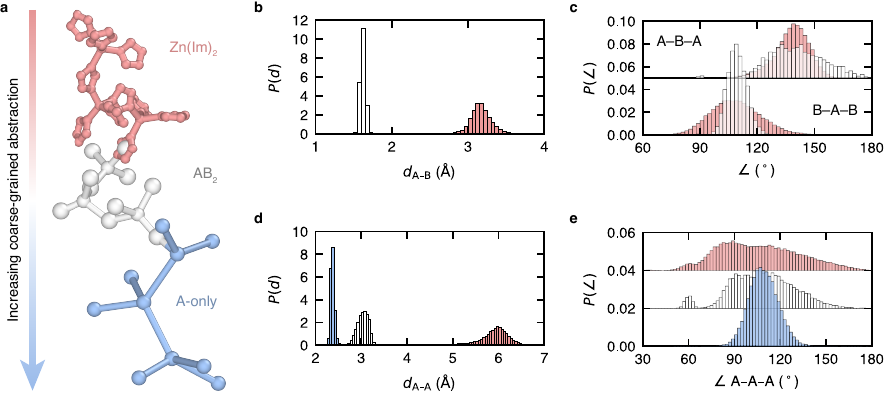}
    \caption{
    \textbf{Comparing bond-length and bond-angle distributions at equivalent levels of coarse-grained abstractions for} {\bf\textit{a}}\textbf{-Si,} {\bf\textit{a}}-\textbf{SiO$_2$, and} {\bf\textit{a}}-\textbf{ZIF.}
    (\textbf{a}) A representative fragment from an AL-HRMC $a$-ZIF model, shown using fully atomistic ({\em top}), AB$_{2}$ coarse-grained (`cg', {\em middle}), and A-only coarse-grained ({\em bottom}) representations, illustrating the increasing degree of abstraction achieved through coarse-graining.
    Comparing the (\textbf{b}) {\sf A}--{\sf B} bond-lengths and (\textbf{c}) {\sf A}--{\sf B}--{\sf A} bond-angles (\textit{top}) and {\sf B}--{\sf A}--{\sf B} bond-angles (\textit{bottom}) for $a$-SiO$_2$ and $a$-ZIF.
    Comparing the (\textbf{d}) {\sf A}--{\sf A} bond-lengths and (\textbf{e}) {\sf A}--{\sf A}--{\sf A} bond-angles for $a$-Si (blue), $a$-SiO$_2$ (white) and $a$-ZIF (red).
    }
    \label{figS6}
\end{figure}

\clearpage

\begin{figure}
    \centering
    \includegraphics[width=\linewidth]{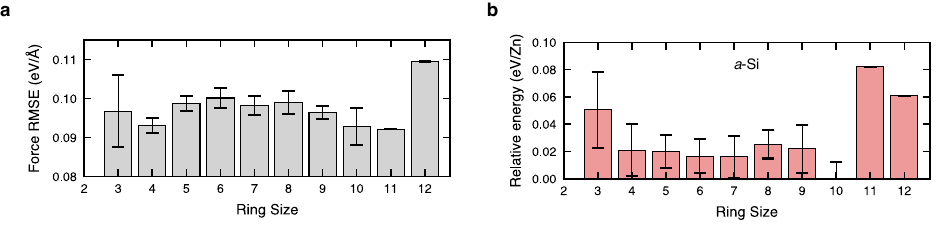}
    \caption{
    \textbf{Validation of the ACE MLIP model for the} {\bf\textit{a}}\textbf{-ZIF topology.}
    (\textbf{a}) Average zinc-node force RMSE as a function of ring size for the topologically disordered ZIF structures in our test set.
    (\textbf{b}) Average zinc local energy as a function of ring size for the same structures.
    Whiskers indicate the standard error of the mean, computed over 10 topologically disordered ZIF structures.
    }
    \label{figS7}
\end{figure}

\begin{figure}
    \centering
    \includegraphics[width=0.5\linewidth]{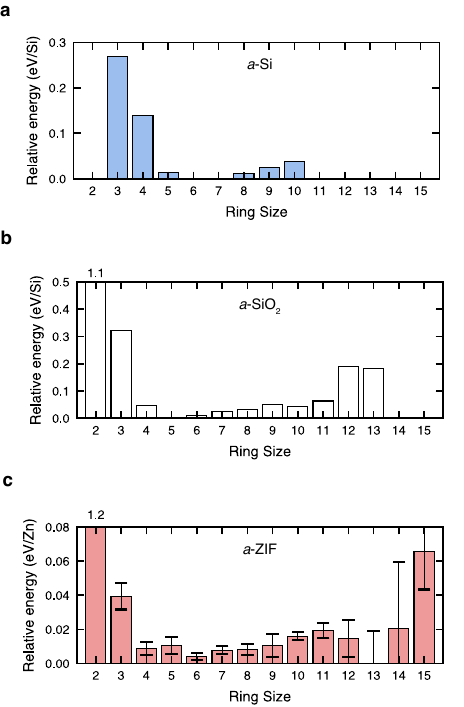}
    \caption{
    \textbf{Comparing local energy distributions as a function of ring size for different tetrahedral networks.}
    The average node local energy as a function of ring size for (\textbf{a}) \textit{a}-Si, (\textbf{b}) \textit{a}-\ce{SiO2}, and (\textbf{c}) \textit{a}-ZIF. Whiskers indicate the standard error of the mean, computed over 10 topologically disordered ZIF structures.
    The `2-membered rings' observed in both \textit{a}-\ce{SiO2} and \textit{a}-ZIF---which consist of two nodes linked by two linkers---exhibit very high energies, with their average energy values indicated above the respective bars.
    }
    \label{figS8}
\end{figure}

\clearpage
\section{Supplementary Tables}

\begin{table}[H]
	\centering
	\caption{Summary of the composition of the ML potential reference dataset.}
	\label{table:database}
	\begin{tabular}{rrrr|cc|cc}
		\hline
		\hline
		& & \multicolumn{2}{c|}{Database Size} & \multicolumn{2}{c}{$\Delta E$ (meV/atom)} & \multicolumn{2}{c}{$F$ (eV/\AA{})} \\
		\cline{3-4} \cline{5-6} \cline{7-8} 
		& & Cells & Atoms & $\mu$ & $\sigma$ & $\mu$ & $\sigma$ \\
		\hline
		\multirow{3}{*}{iter-0} & Crystal & 374 & 101,728 & 96.9 & 79.1 & 10.3 & 9.7 \\
		& Topology & 1,182 & 425,425 & 269.9 & 362.0 & 10.7 & 13.5 \\
		& MOF-FF & 285 & 132,192 & 165.3 & 97.2 & 11.2 & 9.6 \\
		\hline
		GAP iter-1 & MD iter. & 156 & 70,176 & 135.1 & 70.5 & 7.0 & 2.3 \\
		GAP iter-2 & MD iter. & 231 & 94,112 & 118.7 & 60.6 & 6.2 & 2.4 \\
		GAP iter-3 & MD iter. & 282 & 122,400 & 99.9 & 51.4 & 6.3 & 2.6 \\
		GAP iter-4 & MD iter. & 327 & 124,848 & 79.6 & 40.5 & 5.9 & 2.1 \\
		GAP iter-5 & MD iter. & 321 & 141,712 & 143.3 & 86.4 & 7.3 & 3.5 \\
		GAP iter-6 & MD iter. & 96 & 26,112 & 231.8 & 20.1 & 10.5 & 2.0 \\
		\hline
		ACE iter-0 & MD iter. & 525 & 158,304 & 244.7 & 131.5 & 11.0 & 5.7 \\
		\hline
		ACE iter-1 & HRMC iter. & 55 & 29,920 & 174.9 & 132.1 & 8.3 & 10.1 \\
		ACE iter-2 & HRMC iter. & 98 & 53,312 & 102.7 & 62.6 & 6.5 & 3.9 \\
		ACE iter-3 & HRMC iter. & 151 & 82,144 & 74.6 & 24.7 & 3.3 & 0.7 \\
		\hline
		Total & & 4,083 & 1,562,385 & 149.0 & 93.7 & 8.0 & 5.2 \\
		\hline
		\hline
	\end{tabular}
\end{table}

\begin{table}
	\centering
	\caption{
		\label{table:gap-params}
		Final GAP ML potential fitting parameters. 
	}
	\begin{tabular}{l|c|cc|cccc}
		\hline
		\hline
		& 2-body & \multicolumn{2}{c|}{3-body} & \multicolumn{4}{c}{\texttt{TurboSOAP}} \\
		\hline
		&     & Zn    & Im    & Zn    & N  & C    & H \\
		\hline
		$\delta$ (eV)       & 1.0 & 0.087 & 0.087 & 0.053 & 0.053 & 0.053 & 0.053 \\
		$\theta$            & 2.0 & 1.0   & 1.0  \\
		\hline
		$r_{\rm{cut}}$ (Å)  & 6.0 & 2.2   & 1.8   & 6.0   & 6.0   & 6.0  & 6.0   \\
		$r_\Delta$ (Å)      &     &       &       & 0.5   & 0.5   & 0.5  & 0.5   \\
		\hline
		$\sigma_{\rm{at}}$ (Zn) (\AA) & & & & 0.80 & 0.25 & 0.45 &          \\
		$\sigma_{\rm{at}}$ (N) (\AA)  & & & & 0.30 & 0.65 & 0.30 & 0.25     \\
		$\sigma_{\rm{at}}$ (C) (\AA)  & & & & 0.25 & 0.35 & 0.50 & 0.25     \\
		$\sigma_{\rm{at}}$ (H) (\AA)  & & & &      & 0.60 & 0.35 & 0.25     \\
		$n_{\rm{max}}$, $l_{\rm{max}}$   &       &     &       & 8, 4      & 8, 4      & 8, 4     & 8, 4 \\
		$\zeta$                     &       &     &       & 4      & 4      & 4     & 4 \\
		\hline
		Sparsification & Uniform & \multicolumn{2}{c|}{Uniform} & \multicolumn{4}{c}{CUR} \\
		\hline
		$N_{\rm{sparse}}$ (crystalline) &     &       &       & 1,000   & 1,000   & 1,000   &  500  \\
		$N_{\rm{sparse}}$ (topology)    &     &       &       &   500    &  500    &  500   &  250  \\
		$N_{\rm{sparse}}$ (heat)        &     &       &       & 2,500   & 2,000   & 2,000   & 1,500 \\
		$N_{\rm{sparse}}$ (pressure)    &     &       &       & 1,500   & 1,000   & 1,000   &  500  \\
		$N_{\rm{sparse}}$ (liquid)      &     &       &       & 1,000   & 1,000   & 1,000   & 1,000 \\
		$N_{\rm{sparse}}$ (buffer)      &     &       &       & 1,000   & 1,000   & 1,000   & 1,000 \\
		\hline
		$N_{\rm{sparse}}$ (total) & 50$\times$8 & 150$\times4$ & 150$\times2$ & 7,500 & 6,500 & 6,500 & 4,750 \\
		\hline
		\hline
	\end{tabular}
\end{table}

\begin{table}
	\centering
	\caption{GAP regularisation parameters.}
	\label{table:gap-regularisation}
	\begin{tabular}{l|c|c|c}
		\hline
		\hline
		\multirow{2}{*}{Category}  & \multicolumn{3}{c}{Regularisation Parameters} \\
		\cline{2-4}
		& $\sigma_{\rm energy}$ (meV/atom)& $\sigma_{\rm force}$ (eV/\AA)& $\sigma_{\rm stress}$  (eV)\\
		\hline
		Crystalline & 0.2 & 0.002 & 0.002 \\
		Topology & 5 & 0.2 & 0.2 \\
		Heat & 5 & 0.2 & 0.2 \\
		Pressure & 5 & 0.2 & 0.2 \\
		Liquid &  5 & 0.2 & 0.2 \\
		Buffer & 50 & 0.2 & 0.2 \\
		\hline
		\hline
	\end{tabular}
\end{table}

\begin{table}
	\centering
	\caption{Optimised ACE hyperparameters.}
	\label{tab:ace-xpot}
	\begin{tabular}{c|c|c|c|c|c}
		\hline
		\hline
		Model & \texttt{nfunc} & $P$ & $f_{i>2}$ & \texttt{dcut} & \texttt{wlow} \\
		\hline
		1 & 600 & 2 & & 0.01 & 0.86 \\
		2 & 600 & 4 & 0.025, 0.883 & 0.73 & 0.99 \\
		3 & 600 & 6 & 0.975, 0.635, 0.025, 0.540 & 1.00 & 0.85 \\
		\hline
		4 & 900 & 2 & & 0.01 & 0.87 \\
		5 & 900 & 4 & 0.025, 0.760 & 0.11 & 0.76 \\
		\hline
		6 & 1200 & 2 & & 0.97 & 0.85 \\
		\hline
		\hline
	\end{tabular}
\end{table}

\begin{table}
    \centering
    \caption{Quality of fits to experimental total scattering data for RMC and AL-HRMC models.}
    \label{tab:chi2_comparison}
    \begin{tabular}{lcc}
        \hline
        \hline
        \textbf{Model} & $\chi^2(X)$ & $\chi^2(N)$ \\
        \hline
        RMC & 1.99 & 0.68 \\
        AL-HRMC & 0.64 & 0.41 \\
        \hline
        \hline
    \end{tabular}
\end{table}

{

\begin{sidewaystable}
    \centering
    \caption{Vertex symbols and CARVS for crystalline ZIF nets.}
    \label{tab:topology-symbols}
    \footnotesize
    \begin{tabular}{c|c|c|c}
        \hline
        \hline
        Net & Vertex symbol & All-rings vertex symbol & CARVS \\
        \hline
        \textbf{cag} & [4.6$_2$.6.6.6.6] & [(4,8).(6$_2$,8).(6,8$_4$).(6,8$_4$).(6,8$_3$).(6,8$_3$)] & \{ 4.6$_6$.8$_{16}$ \} \\
        \hline
        \textbf{zni} & [4.6.6.6$_3$.6$_2$.12$_{40}$] & [(4,6$_2$,14$_{10}$).(6,14$_{48}$).(6,12$_{40}$,14$_{31}$).(6$_{3}$,12$_{20}$).(6$_{2}$,12$_{20}$,14$_{10}$).(12$_{40}$,14$_{41}$)] & \{ 4.6$_{9}$.12$_{120}$.14$_{140}$ \} \\
        \hline
        \textbf{sod} & [4.4.6.6.6.6] & [(4,12$_4$).(4,12$_4$).(6,12$_6$).(6,12$_6$).(6,12$_6$).(6,12$_6$)] & \{ 4$_2$.6$_4$.12$_{32}$ \} \\
        \hline
        \multirow{2}{*}{\textbf{zec}} & 1$\times$ [6.6.6$_2$.6$_2$.6$_2$.6$_2$] & 1$\times$ [(6,10).(6,10).(6$_2$,8$_2$,10,12$_4$.(6$_2$,8$_2$,10,12$_4$.6$_2$.6$_2$] & \multirow{2}{*}{ \{4$_{1.6}$.6$_{6}$.8$_{3.2}$.10$_4$.12$_{9.6}$\}$_{\sigma=2.3}$ } \\
                                      & 4$\times$ [4.6.4.6$_3$.6.10$_2$]        & 4$\times$ [(4,8,12).(6,10$_2$,12$_3$.(4,12$_2$.(6$_3$,8).(6,8,12).(10$_2$,12$_3$] & \\
        \hline
        \multirow{4}{*}{\textbf{coi}} & 1$\times$ [4.8$_{3}$.4.8$_{4}$.8$_{9}$.10]      & 1$\times$ [(4,10$_{6}$,14$_{7}$).(8$_{3}$,10$_{17}$,14$_{20}$,16).(4,14$_{7}$.(8$_{4}$,10$_{5}$,14$_{27}$,16$_{3}$.(8$_{9}$,10$_{4}$,14$_{2}$.(10,14$_{24}$,16$_{4}$] & \multirow{4}{*}{\{ 4$_{2.5}$.8$_{14}$.10$_{32.5}$.14$_{63}$.16$_{8}$ \} $_{\sigma=18.9}$ } \\
                                      & 1$\times$ [4.8$_{5}$.4.8$_{7}$.8$_{4}$.10$_{3}$]& 1$\times$ [(4,8$_{2}$,10$_{5}$.(8$_{5}$,10$_{21}$,14$_{20}$,16$_{2}$.(4,8$_{3}$.(8$_{7}$,14$_{13}$,16$_{2}$.(8$_{4}$,10$_{5}$,14$_{16}$.(10$_{3}$,14$_{5}$,16$_{3}$] & \\
                                      & 1$\times$ [4.4.4.8$_{7}$.8.8$_{2}$] & 1$\times$ [(4,10$_{6}$,14$_{7}$.(4,14$_{7}$.(4,8$_{2}$,10$_{4}$,16).(8$_{7}$,10$_{4}$,14$_{4}$,16$_{3}$.(8,10$_{14}$,14$_{14}$,16$_{2}$.(8$_{2}$,10$_{4}$,14$_{7}$,16$_{2}$] & \\
                                      & 1$\times$ [4.4.4.10$_{5}$.8.8] & 1$\times$ [(4,8$_{3}$,16).(4,8$_{2}$,10$_{5}$.(4,10$_{6}$,16$_{2}$.(10$_{5}$,14$_{37}$,16$_{5}$.(8,10$_{7}$,14$_{21}$,16).(8,10$_{8}$,14$_{14}$] & \\
        \hline
        \multirow{5}{*}{\textbf{nog}} & 1$\times$ [5.5.5.6.8.8$_6$] & 1$\times$ [(5,6,12).(5,12).(5,12).(6,12).(8,12$_4$).8$_6$] & \multirow{5}{*}{ \{5$_3$.6$_{1.2}$.8$_{8.8}$.12$_{7.2}$\}$_{\sigma=4.0}$ } \\
                                      & 1$\times$ [4.6.5.5.5.8$_3$] & 1$\times$ [4.(6,12$_2$).(5,8$_3$).(5,8$_3$).(5,8$_3$).8$_3$] & \\
                                      & 1$\times$ [5.5.5.5.8.8$_2$] & 1$\times$ [(5,12).(5,8$_2$,12).(5,12).(5,8$_2$,12).(8,12$_4$.(8$_2$,12$_2$] & \\
                                      & 1$\times$ [4.8$_3$.5.5$_2$.8$_2$.8$_2$] & 1$\times$ [(4,8$_4$.(8$_3$,12$_6$.5.5$_2$.8$_2$.8$_2$] & \\
                                      & 1$\times$ [5.5.6.8.6.8$_2$] & 1$\times$ [(5,6,12).(5,8$_2$,12).(6,8$_2$,12$_2$.(8,12$_4$.(6,12).(8$_2$,12)] & \\                                                 
        \hline
        \hline
    \end{tabular}
\end{sidewaystable}
}